\documentclass[letterpaper, showkeys,onecolumn,11pt,unpublished]{quantumarticle}
\pdfoutput=1
\usepackage[utf8]{inputenc}
\usepackage[english]{babel}
\usepackage[T1]{fontenc}
\usepackage{amsmath}

\usepackage{tikz}
\usepackage{lipsum}

\usepackage[
    colorlinks=true,
    linkcolor=blue,
    urlcolor=blue,
    citecolor=blue]{hyperref}
    
\usepackage{lineno}
\usepackage{dcolumn}
\usepackage{bm}
\usepackage{mathtools}
\usepackage{braket}
\usepackage{physics}
\usepackage{afterpage}
\usepackage{nicefrac}
\usepackage{xcolor}
\usepackage{amsfonts}
\usepackage{amssymb}
\usepackage{comment}
\usepackage{stfloats}
\usepackage{float}

\usepackage[numbers,sort&compress]{natbib}

\begin{document}

\title{Mirrorless lasing: a theoretical perspective}

\author{Aneesh Ramaswamy}
\affiliation{Department of Physics, Stevens Institute of Technology, Hoboken, NJ 07030, USA }

\author{Jabir Chathanathil}
\affiliation{DEVCOM Army Research Laboratory, Adelphi, MD 20783, USA}

\author{Dimitra Kanta}
\affiliation{Helmholtz Institute Mainz, Johannes Gutenberg University, 55099 Mainz, Germany}

\author{Emmanuel Klinger}
\affiliation{Helmholtz Institute Mainz, Johannes Gutenberg University, 55099 Mainz, Germany}
\affiliation{Universit\'e de Franche-Comt\'e, SupMicroTech-ENSMM, UMR 6174 CNRS, Institut FEMTO-ST, 25000 Besan\c{c}on, France}

\author{Aram Papoyan}
\affiliation{Institute for Physical Research, National Academy of Sciences of Armenia, 0203 Ashtarak-2, Armenia}

\author{Svetlana Shmavonyan}
\affiliation{Institute for Physical Research, National Academy of Sciences of Armenia, 0203 Ashtarak-2, Armenia}

\author{Aleksandr Khanbekyan}
\affiliation{Institute for Physical Research, National Academy of Sciences of Armenia, 0203 Ashtarak-2, Armenia}

\author{Arne Wickenbrock}
\affiliation{Helmholtz Institute Mainz, Johannes Gutenberg University, 55099 Mainz, Germany}

\author{Dmitry Budker}
\affiliation{Helmholtz Institute Mainz, Johannes Gutenberg University, 55099 Mainz, Germany}
\affiliation{Department of Physics, University of California, Berkeley, CA 94720, USA}

\author{Svetlana A. Malinovskaya}
\affiliation{Department of Physics, Stevens Institute of Technology, Hoboken, NJ 07030, USA }
\affiliation{Helmholtz Institute Mainz, Johannes Gutenberg University, 55099 Mainz, Germany}

\maketitle

\begin{abstract}

Mirrorless lasing has been a topic of particular interest for about a decade due to promising new horizons for quantum  science and applications. In this work, we review first-principles theory that describes this phenomenon, and discuss degenerate mirrorless lasing in a  vapor of Rb atoms, the mechanisms of amplification of light generated in the medium with population inversion between magnetic sublevels 
within the $D_2$ line, and challenges associated with experimental realization.

\end{abstract}

\keywords{Quantum Optics, Stimulated emission processes, Mirrorless lasing, Amplified spontaneous emission, Alkali vapors}

\section{\label{sec:level1}Introduction}

Over sixty years of existence, lasers (Light Amplification by Stimulated Emission 
of Radiation) have played a significant role in many areas of scientific research, industry and defense \cite{Hecht2010}, continuously growing as new laser technologies are developed.
There are three principal components usually attributed to a laser: a gain medium, a pumping process and a feedback loop \cite{Siegman1986}. Lasing usually requires pumping the medium to a state of population inversion, although lasing without apparent inversion can occur in the case where quantum coherence is induced between lower levels \cite{Scully1992}. There is a debate over whether lasing always requires a feedback loop. Lasing is often distinguished from processes such as Amplified Spontaneous Emission\,(ASE), Superradiance\,(SR) and Superflouorescence\,(SF) \cite{Dudok2013,Malcuit87}. Conventional lasers usually incorporate an optical resonator setup where mirrors are used to have light amplified over several round trips in the gain medium \cite{Siegman1986}. In mirrorless lasing setups involving a feedback loop,
the gain medium takes the role of the resonator -- usually through multiple scattering processes \cite{Dudok2013} -- in systems with a range of disorder, including random lasers \cite{Wiersma2008} and distributed-feedback systems \cite{Schilke2011}. Lasing in gain medium with a nonresonant feedback loop provided by the same medium was considered by Letokhov \cite{Letokhov1968} who gave a theoretical treatment concerning light in the diffusive regime, see, for example, Ref.\,\cite{Wiersma1996}. Disordered random media provide coherent feedback loops that generate lasing as seen in disordered ZnO nanoparticles in polycrystalline films \cite{Cao2003} and quantum-dot-doped liquid crystals\,\cite{Wang2020}. In random lasers, feedback is either resonant (phase sensitive,  i.e. coherent) or non-resonant (frequency and phase independent, i.e. incoherent)\,\cite{Cao2003}. 


In this work, we adopt a broad definition of mirrorless lasing as directed monochromatic emission from an ensemble of atoms or molecules excited with pump laser light. The treatment of feedback loops arising from multiple scattering is not considered here. 
There is great interest in the phenomenon of mirrorless lasing in atomic vapors with theoretical \cite{Zhu1992,Scully1992,Fearn1992,Guerin2008,Labeyrie1999,Binninger2019,Rosato2013,Kupriyanov2003,Gremaud2006}, and experimental investigations \cite{Baudouin,Akulshin2018, Papoyan2019,gazazyan,Cherroret2021,movsisyan} conducted in both cold atoms and hot vapors. The problem of mirrorless lasing in atomic gases can be split into two: $(i)$ the problem of the gain mechanism \cite{Guerin2008,Baudouin} and $(ii)$ the problem of multiple-scattering feedback mechanism \cite{Binninger2019,Rosato2013,Dudok2013}. Mechanisms involving population inversion include Mollow gain (using a near-resonant strong pumping field), Raman gain (by driving transitions using off-resonance fields between two non-degenerate ground states, usually Zeeman or hyperfine levels), and parametric gain (using degenerate four-wave mixing (d-FWM) by way of counter-propagating pump fields) \cite{Guerin2008,Baudouin}. 

Experiments in alkali metal vapor have shown the presence of gain through the phenomenon of amplified spontaneous emission (ASE), see e.g. \cite{Zhdanov2021} and Refs. therein. The combination of ASE and d-FWM at above-threshold scatterer densities is suggested as a likely mechanism \cite{Papoyan2019,Akulshin2018}.

The present work covers theoretical fundamentals of the lasing mechanisms of a gas of atoms in free space,  focuses on the phenomenon of amplification of spontaneous emission and discusses degenerate mirrorless lasing from a gas of alkali atoms with magnetically degenerate hyperfine states. 
 We elucidate the mechanisms for mirrorless lasing in both forward and backward direction with respect to the pump laser beam and show that such process is possible even in the degenerate case of the directed light being of the same frequency as the pump. We reveal the role of population inversion among degenerate magnetic sublevels of the hyperfine manifolds on the light amplification. The paper is organized as follows. In Sec.\,\ref{sec:formalism} we present a first-principles theory describing mirrorless lasing and mechanisms of light amplification in a gas of atoms. In Sec.\,\ref{sec:mirrorlessLasing} we present a study of degenerate mirrorless lasing in rubidium vapor using a semiclassical approach.  Finally in Sec.\,\ref{sec:experiments} we review the current status of experimental investigation of degenerate mirrorless lasing followed by a summary.

\section{First-principles formalism}\label{sec:formalism}
We start with a microscopic approach for the problem of mirrorless lasing in a gas of atoms. A system of N multilevel atoms located in free space and driven by a classical pump field with envelope $E_p(t)$ and frequency $\omega_p$ is considered. Each atom $A_i$ has momentum $\boldsymbol{\mathrm{p}}_i$, center-of-mass (COM) position $\boldsymbol{\mathrm{R}}_i$ and relative position $\boldsymbol{\mathrm{r}}_i$. We define the central potentials $V_i(R_i,r_i)$ and the atom-atom interactions mainly composed of the potentials, $W_{ij}(R_i,R_j,r_i)$, introduced by the repulsion between the electrons of atoms $A_i$ and $A_j$ respectively. 
 The separation distance between any pair of atoms is in general much greater than the transition wavelength $\lambda_p$. 
Each atom $A_i$ is travelling with COM velocity $\boldsymbol{\mathrm{v}}_i$, such that transition frequencies and dipole moments are Doppler shifted.
The dipole approximation is used such that the interaction with the classical field driving transition $l$ for the $i$-th atom is given by $-\mu_{il} E_p(t)\cos(\omega_pt)(\sigma_{il}^{+}+\sigma_{il}^{-})$. The vacuum interaction is given by $g_{\vec{k};il}^{(\mu)}a_{\vec{k}}^{(\mu)}\sigma_{il}^{+}+\mathrm{H.c}$. We define the Hamiltonians:
\begin{equation}
    H_A=\sum_{i}\left(\dfrac{\boldsymbol{\mathrm{p}}_i^2}{2m}+V_i(\vec{R}_i,\vec{r}_i)+\sum_{j\neq i} W_{ij}(\vec{R}_i,\vec{R}_j,\vec{r}_i)\right),
\end{equation}
\begin{equation}
    H_F=\dfrac{1}{8\pi}\int d^3r\text{}\left(\abs{\boldsymbol{\mathrm{E}}}^2+\abs{\boldsymbol{\mathrm{B}}}^2\right),
\end{equation}
\begin{equation}\label{haf}
    H_{AL}=-\sum_{il}\mu_{il} E_p(\vec{R}_i,t)\cos(\omega_pt)\left[\sigma_{l}^{+}(\vec{R}_i)+\sigma_{l}^{-}(\vec{R}_i)\right],
\end{equation}
\begin{equation}
    H_{AF}=-\sum_{il}\int d^3k\text{ }g_{\vec{k};il}^{(\mu)}(\vec{R}_i)a_{\vec{k}}^{(\mu)}\sigma_{l}^{+}(\vec{R}_i)+\mathrm{H.c.}
\end{equation}

From the above Hamiltonians, we determine the atomic density matrix equations and the Heisenberg equation of motion for the photon operator $\expval{a_{\vec{k}}^{(\mu)\dag}a_{\vec{k}}^{(\mu)}}$ \cite{Gremaud2006}. We ignore the contribution from the exchange potential terms and we use the operator $U_L(t,t_1)=\exp{-i\int_{t_1}^{t}H_{AL,I}(t^{\prime})}\exp{-i\left(H_{A}+H_F\right)t}$ to transform the system Hamiltonian to the field interaction reference frame  where the atomic propagators are dressed with the pump field. This is to include the case where the system is continuously driven, as opposed to where the system is optically pumped and then left alone. The latter approach is often used in solid-state lasing gain media, where the presence of non-radiative emissions and quenching decrease the material radiative lifetimes \cite{Burshtein2010}. We note that there are separate, often competing \cite{Boyd1987}, mechanisms for gain. A seed pulse can be continuously amplified by the system or ASE can be generated. We use the standard projection operator techniques to derive the time non-local Liouville-von-Neumann equation \cite{Agarwal1974}: 
\begin{align}\label{Atomb}
\begin{split}
    \dfrac{d}{dt}\rho_{A,I^{\prime}}=&-i\left(\sum_{il}\text{ }g_{\vec{k}l}^{(\mu)}(\vec{R_i})\comm{\sigma_{l,I^{\prime}}^+(\vec{R}_i,t)}{\Omega^{0+}(\vec{R_i},t)}+\text{H.c}\right)\\
    &+\sum_{ilm}\int_{t_0}^{t}dt^{\prime}\left(\Gamma_{lm}(\vec{R_i},t,t^{\prime})\comm{\hat{\sigma}^{+}_{l,I^{\prime}}(\vec{R_i},t)}{\hat{\sigma}^{-}_{m,I^{\prime}}(\vec{R_i},t^{\prime})\rho_{A,I^{\prime}}(t^{\prime})}\right.\\
    &~~\left.-\Gamma_{lm}(\vec{R_i},t,t^{\prime})^{*}\comm{\hat{\sigma}^{-}_{l,I^{\prime}}(\vec{R_i},t)}{\rho_{A,I^{\prime}}(t^{\prime})\hat{\sigma}^{+}_{m,I^{\prime}}(\vec{R_i},t^{\prime})}\right)\\
    &-\sum_{il}\sum_{jm}\dfrac{1}{\hbar^2}\int_{t_0}^{t}dt^{\prime}\biggl(\comm{\hat{\sigma}^{+}_{l,I^{\prime}}(\vec{R_i},t)}{\hat{\sigma}^{-}_{m,I^{\prime}}(\vec{R_j},t^{\prime})\xi^{(\mu^{\prime}\mu)}_{m,l}(\vec{R_j},\vec{R_i},t^{\prime},t)}\\
    &~~\left.-\comm{\hat{\sigma}^{+}_{l,I^{\prime}}(\vec{R_i},t)}{\xi^{(\mu^{\prime}\mu)}_{m,l}(\vec{R_j},\vec{R_i},t^{\prime},t)^{*}\hat{\sigma}^{-}_{m,I^{\prime}}(\vec{R_j},t^{\prime})}\right.\,+\text{H.c.}\biggl),
\end{split}
\end{align}
where 
$\Omega^{0+}(\vec{R_i},t)$ is the vacuum Rabi field operator
\begin{equation}
    \Omega^{0+}(\vec{R_i},t)=\dfrac{1}{\hbar}\int d^3k \expval{a_{\vec{k}}^{(\mu)}}_F(t_0)e^{-i\omega_kt}, 
\end{equation}
$\Gamma_{lm}(\vec{R_i},t,t^{\prime})$ is the time-dependent correlator for the decay
\begin{equation}
    \Gamma^{(\mu)}_{lm}(\vec{R_i},t,t^{\prime})=\int d^3k\text{ }g_{\vec{k};l}^{(\mu)}(\vec{R_i},t)g_{\vec{k};m}^{(\mu)*}(\vec{R_i},t^{\prime}),
\end{equation}
and $\xi^{(\mu^{\prime}\mu)}_{m,l}(\vec{R_j},\vec{R_i},t^{\prime},t)$ represents stimulated emission processes driven by the quantum light fields
\begin{equation}\label{Efldterm}
    \xi^{(\mu^{\prime}\mu)}_{m,l}(\vec{R_j},\vec{R_i},t^{\prime},t)=\expval{\left\lbrace\vec{d}^{*}_m\cdot \vec{E}^{(\mu^{\prime})-}(\vec{R_j},t^{\prime})\right\rbrace\left\lbrace\vec{d}_l\cdot \vec{E}^{(\mu)+}(\vec{R_i},t)\right\rbrace}_{F}.
\end{equation}

The first term in Eq.\,\eqref{Atomb} is the Langevin force that depends on vacuum fluctuations. The second term is the spontaneous decay contribution and $\hat{\sigma}^{\pm}_{l}(\vec{R_i},t)$ is the rotated dipole operator. Note that the time and position dependent exponential is included 
into the coupling rates. The final term is the stimulated absorption/emission term for a single atom, ($i=j$), and two-atom  events, ($i\not=j$). The former provides a correction to spontaneous emission due to a non-zero quantum radiation field.  
We note that there is a dependence on the expectation value $\expval{a_{\vec{k}}^{(\mu)\dag *}a_{\vec{k}^{\prime}}^{(\mu^{\prime})}}_{F}(t^{\prime})$. In the case where the density matrix is separable, and $\vec{k}=\vec{k}^{\prime}$ and $\mu=\mu^{\prime}$, we get the product of the usual photon number operator and the atomic density matrix. When any of the last two conditions are not met, the final term represents a photonic coherence between two modes.
The Heisenberg equations of motion for the photon number operator can be simply defined as 
\begin{equation}\label{Radb}
\begin{split}
    \left[\dfrac{1}{c}\dfrac{d}{dt}+i(\omega_k-\omega_{k^{\prime}})\right]\expval{a_{\vec{k},I^{\prime}}^{(\mu)\dag}(t)a_{\vec{k}^{\prime},I^{\prime}}^{(\mu^{\prime})}(t)}(t)=S_{\text{s}}(t)+\mathcal{R}_{\text{s}}\left[\expval{a_{\vec{k}/\vec{q},I^{\prime}}^{(\mu)\dag}a_{\vec{q}/\vec{k}^{\prime},I^{\prime}}^{(\mu^{\prime})}}(t)\right],
\end{split}
\end{equation}
where $S_S(t)$ is the source term due to atomic spontaneous decay and $\mathcal{R}_{\text{s}}\left[\expval{a_{\vec{k}/\vec{q},I^{\prime}}^{(\mu)\dag}a_{\vec{q}/\vec{k}^{\prime},I^{\prime}}^{(\mu^{\prime})}}(t)\right]$ represents scattering contributions involving other modes. Equations \eqref{Atomb} and \eqref{Radb} are the microscopic equations required to model the time and space dependent atomic density matrix and quantum field equations. For optically thick systems with high disorder and/or non-ultracold temperatures,  statistical averaging of the equations, including averaging over the position density and velocity distributions of atoms, and phase-space transforms are required to obtain macroscopic equations of motion \cite{Rosato2011}.

\subsection{The probe field gain}
The most common method for generating an intense monochromatic beam is to amplify a weak probe pulse, $E_{pr}(t)$, using a medium which has undergone population inversion, in the dressed state basis, with respect to a transition having frequency close to the probe frequency \cite{Agarwal1991}. Additional methods involve using coherent fields to generate quantum interference between stimulated emission and absorption pathways to create gain without population inversion \cite{Mompart2000, Scully1992, Baudouin, Guerin2008}. The probe detuning, $\Delta_{pr}$, the magnetic sub-level degeneracy, the presence of population inversion and the propagation direction of the pump beam(s) determine which amplification  mechanism dominates.

A Hamiltonian analogous to that in Eq.\,\eqref{haf} is used to model the atom-probe field interaction, which we call $H_{pr}$. We derive the rate equations for the probe field $E_{pr}$. Using the steady state absorption rate of quanta for a classical field \cite{Mollow1972}, we write
\begin{equation}
    \dfrac{\partial }{\partial t}E_{pr}(\vec{R},t)=n(\vec{R})\expval{-i\comm{H_{pr;I}(t)}{\rho_I(t)}}\Delta(\vec{R},\vec{r}_0)\,,
\end{equation}
where $n$ is the density of atoms in the medium, and $I$ denotes the field interaction picture associated with transformation $U_0(t,t_1)=\exp{-i\left(H_{A}+H_F\right)t}$. The terms in $H_{AL;I}(t)$ are restricted to atoms within a volume $\Delta(\vec{R},\vec{r}_0)$ of radius $\vec{r}_0$ from the center $\vec{R}$. 

To the first order in perturbation theory, the steady state of a driven atom results in absorption spectrum $g_A(\omega)$ of a weak probe field of mode $M$ and frequency $\omega$ \cite{Mollow1972}
\begin{align}\label{spectraltt}
    g_A(\omega)=\int_{0}^{\infty}d\tau\text{ } e^{-i\omega t} \comm{d_I^-(\tau)}{d_I^+(0)}\,,
\end{align}
where $d_I^+(\tau)$ is the weighted sum of all dipole operators that correspond to transitions involving radiation of mode $M$:
\begin{align}\label{dipel}
    d_I^+(\tau)=\sum_j \left(\hat{\varepsilon}\cdot\vec{\mu}_j\right)\sigma_{j,I}^+(\tau).
\end{align}

 The averaged single-atom absorption rate, $\alpha(\omega,t)$, of quanta of mode $M$ and energy $\hbar\omega$ from a weak probe field with slowly varying amplitude $E_{pr}(t)$ and frequency $\omega$ according to  \cite{Mollow1972} is
\begin{align}
\alpha(\omega,t)=\expval{g_A(\omega)}\left|\dfrac{\left(\hat{e}_p\cdot\hat{d_I}\right)E_{pr}(t)}{\sqrt{2}\hbar}\right|^2\,.
\end{align}
Understanding of the scattering processes that contribute to stimulated emission of the probe field in the presence of strong driving pump fields requires a dressed state approach. In strong fields, the spectral state composition of the atoms is understood in terms of multi-photon dressed states $\ket{\lambda_i,m}$ with energies $E_i=\lambda_i+m\omega_p$. In this case, the unperturbed atomic dynamics at any time consists of ladder-type transitions between different $m$-number states. Therefore, contributions to the stimulated emission of probe field photons must take into account these background ladder transitions \cite{Chou2010}. 

The transition amplitudes for multi-photon processes can be found by looking at the diagrammatic expansion of the self-energy $\Sigma(t,t_0)$ \cite{Mueller2017}. This follows from the below form of the Liouville-von-Neumann equation:
\begin{equation}
    \dot{\rho}_I(t)=\int_{t_0}^{t}dt^{\prime}{\rho}_I(t^{\prime})\Sigma(t^{\prime},t).
\end{equation}
For example, the sum of transition amplitudes for all three-photon processes that involve absorption of two pump photons and emission of one single probe photon is given by summing over three-vertex Keldysh diagrams with vertices $(t_1, \sigma_j^+E_P^{-})$, $(t_2, \sigma_l^-E_{pr}^{+})$, $(t, \sigma_j^+E_P^{-})$. The vertices can be on the forward or backwards branch of the density matrix, and we sum over all permutations of the vertices.

\begin{figure}[h!]\centering
    \includegraphics[scale=0.35]{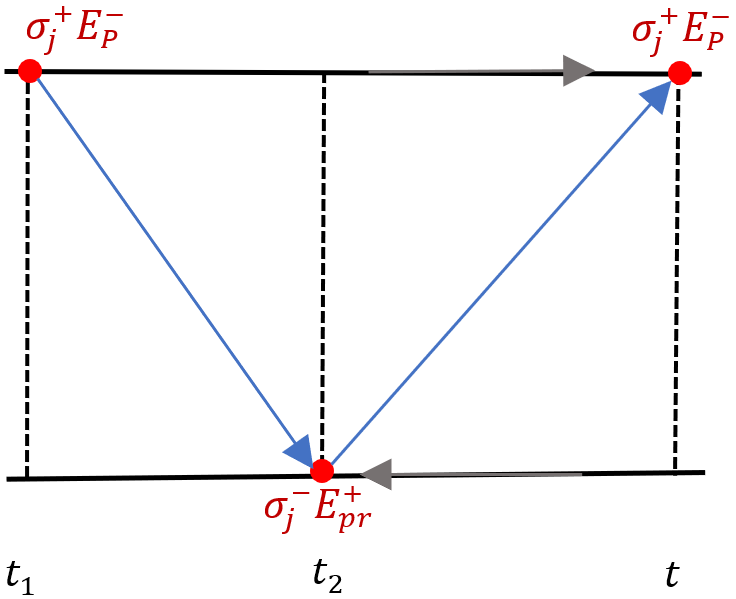}
    \caption{A diagram representing the three-photon scattering process.}
    \label{fig: MollowRb}
\end{figure}
For the three-photon scattering process involving absorption of two pump photons and emission of one single probe photon, the contribution to the evolution of the atomic density matrix is given by
\begin{equation}
\begin{split}
    \delta^{(3)}\rho_{A,I}(t_1)&=\dfrac{E_P^2E_{pr}}{\hbar^3}\int_{0}^{t}\int_{0}^{t_2}dt_1dt_2 e^{-i\left((\Delta_p)(t+t_1)-\Delta_{pr}t_2\right)}\\
    &~\times\Big[d_{P}^{+}U(t,t_2)d_{pr}^{-}U(t_2,t_1)d_{P}^{+}\rho_{A,I}(t_1)U^{\dag}(t,t_1)\\
    &~+ \,U(t,t_1)\rho_{A,I}(t_1)d_{P}^{+}(t_2)U^{\dag}(t_2,t_1)d_{pr}^{-}U^{\dag}(t,t_2)d_{P}^{+}(t)+...\Big],
\end{split}
\end{equation}
where suspension points refer to all other possible time and branch (forward or reverse of the density matrix) orderings of the three vertices, see Fig.\,\ref{fig: MollowRb}.
The contribution in the steady state limit is determined by substituting the steady state expression $\bar{\rho}_{A,I;s}=\bar{\rho}_{A,I;0}+\sum_j \bar{\rho}_{A,I;j}e^{-i\nu_jt}$. To look at the contribution for a specific transition, we compute the change in energy of the probe field, $\expval{H_p\hat{X}}$, for the particular transition described by the string of operators, $\hat{X}$.

In the following subsections, we described possible gain mechanisms using the example of a two-level degenerate system modeling the $F=2\rightarrow F=3$ component of $^{87}$Rb $D_2$ transition.

\subsection{Mollow gain}
\begin{figure}[h!]\centering
    \includegraphics[scale=0.2]{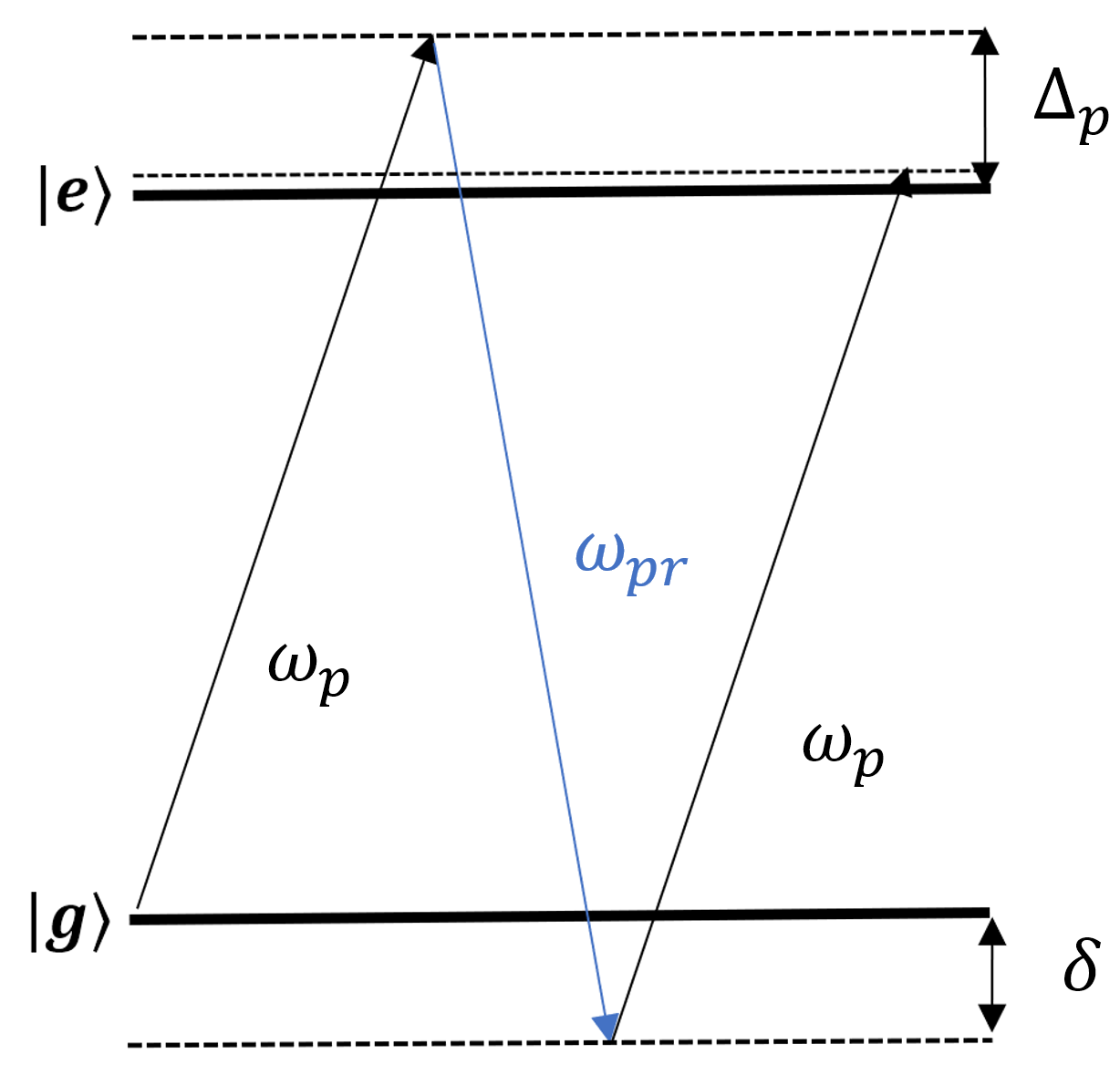}
    \caption{Mollow gain in a two-level system (TLS). The pump field 
 ($\omega_{pr}$), with detuning $\Delta_p$, pumps atoms into the higher energy dressed state. Depending on the sign of $\Delta_p$, the probe field with frequency $\omega_{pr}$ and two-photon detuning $\delta$ is either amplified or reduced due to a three-photon scattering process. The probe field polarization is parallel to the polarization of the pump.}\label{Mollow}
\end{figure}

\begin{figure}[h!]\centering
    \includegraphics[scale=0.25]{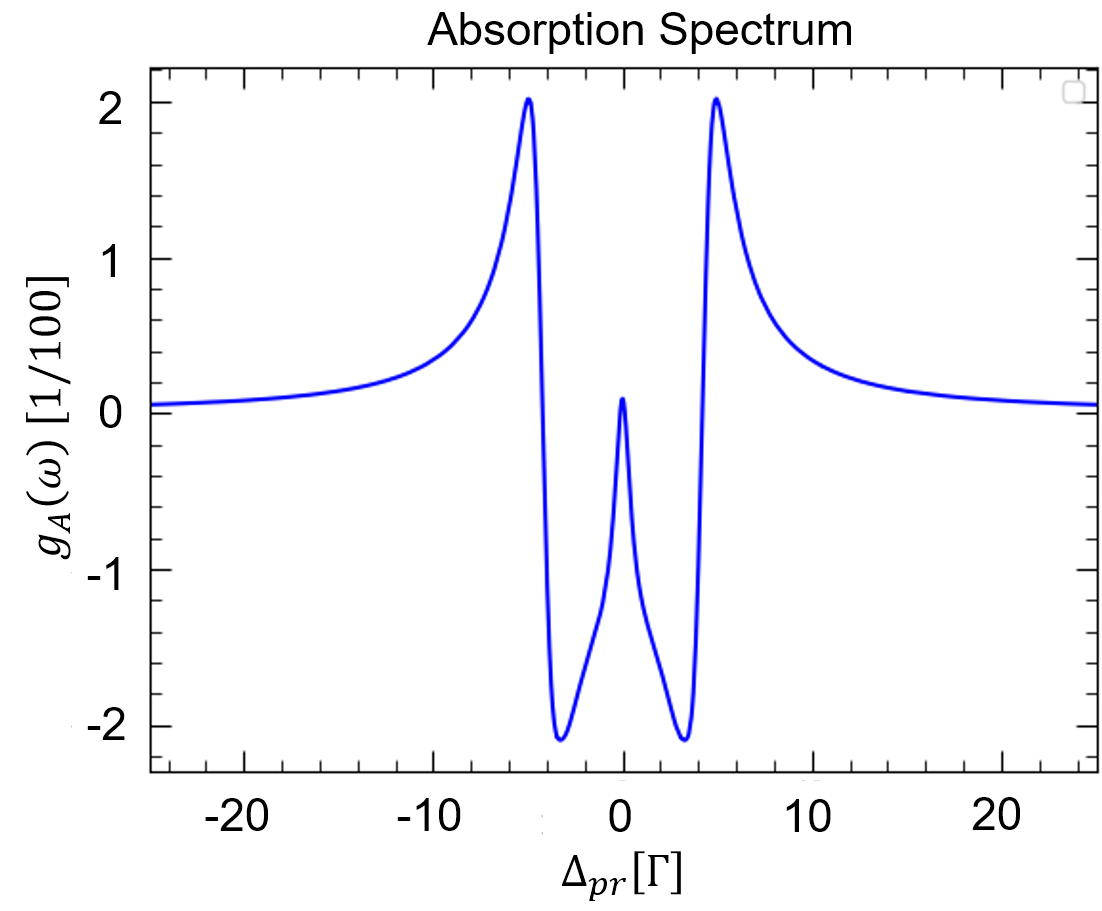}
    \caption{Absorption spectrum of light with parallel (to pump) polarization for the resonantly driven $^{87}$Rb  system with parameters $\Gamma=1.0$, $\Omega_{p}=4\,\Gamma$, $\Delta_p=0$.}
    \label{fig: MollowRb1}
\end{figure}

\begin{figure}[h!]\centering
    \includegraphics[scale=0.25]{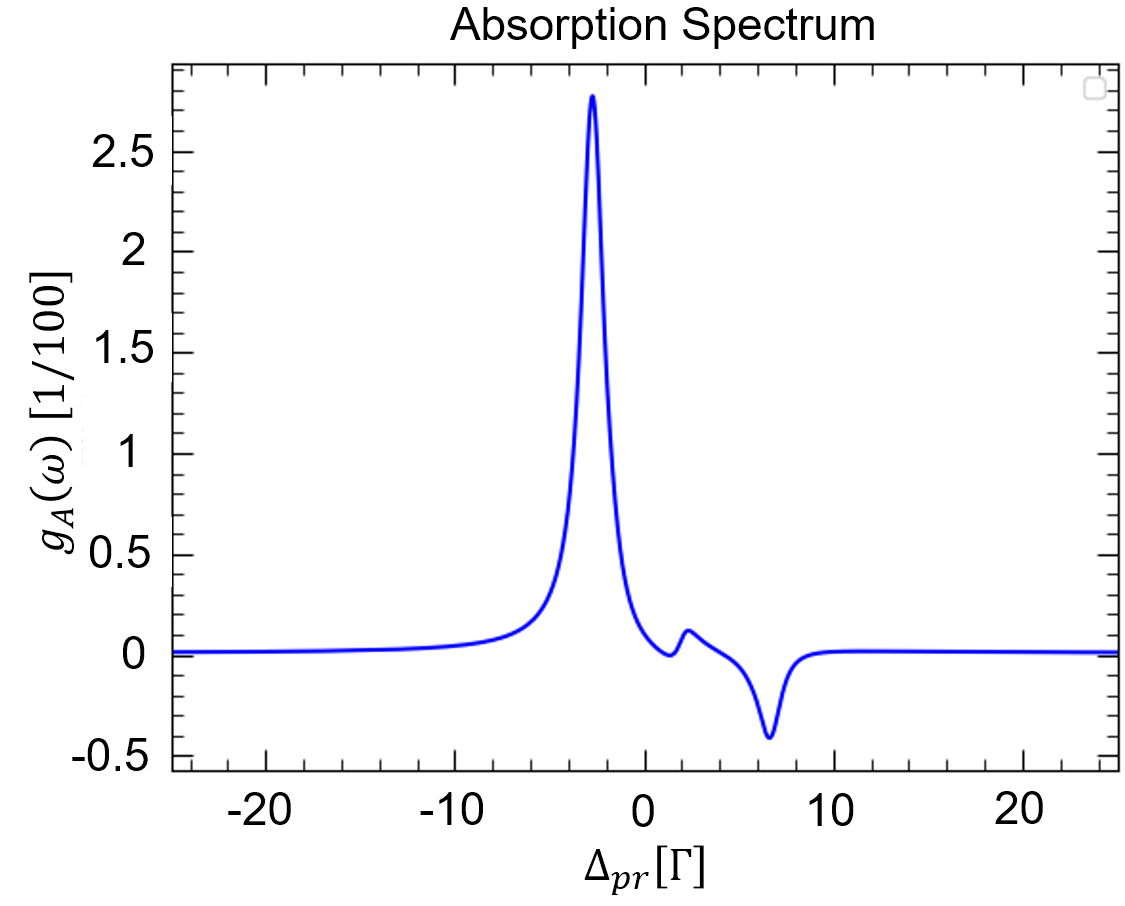}
    \caption{Absorption spectrum of light with parallel (to pump) polarization for the off-resonantly driven $^{87}$Rb system with parameters $\Gamma=1.0$, $\Omega_{p}=4\Gamma$, $\Delta_p=2\Gamma$}
    \label{fig: MollowRb2}
\end{figure}

\begin{figure}[h!]\centering
    \includegraphics[scale=0.25]{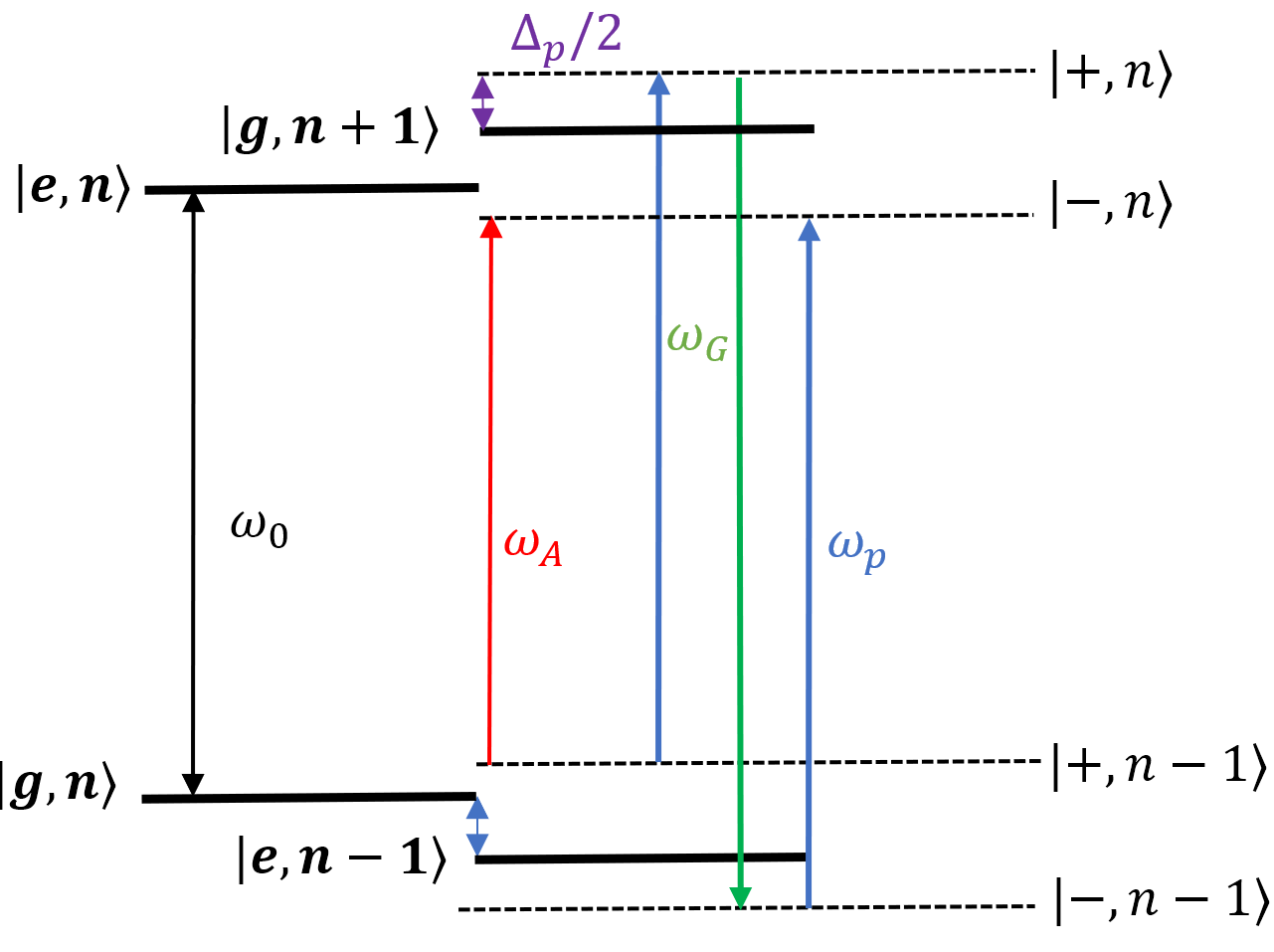}
    \caption{Atom-photon dressed state picture of the TLS in the case when  the pump pulse is off resonance; $\omega_A$ corresponds to an absorption feature that appears with non-resonant driving; $\omega_G=2\omega_{pr}-\omega_A$ is the probe frequency that corresponds to the three-photon scattering
    amplification side-band.}
    \label{fig: Keldysh3sc}
\end{figure}

In the case where the probe field is near resonance and a single pump field is used, Mollow gain \cite{Mollow1972} dominates in which the gain mechanism is due to a three-photon process involving the absorption of two pump photons and stimulated emission of a probe photon \cite{Baudouin}, see Fig.\,\ref{Mollow}. 
Population inversion in the dressed state basis has occurred here and resulted in gain/loss in the sidebands of the absorption spectrum about the resonance peak \cite{Chal/upczak1994}. 

Figure \ref{fig: MollowRb1} shows the absorption spectrum for the case of resonant driving field. Note that, in this work, frequency parameters (such as $\Delta_{p,pr}$) are expressed in units of $\Gamma$, the excited state decay rate, and time parameters in the units of $1/\Gamma$. The feature seen at $\Delta_p=0$ corresponds to an interference between multiple two-photon processes. The dominant resonant contribution is the two-photon Rayleigh scattering process involving an absorption of the probe photon and spontaneous emission of a photon having  the same frequency \cite{Grynberg1993}. Additionally there are processes involving absorption of a pump photon and emission of a probe photon and vice versa. 
The greatest gain components are observed at the sidebands centered at the dressed state frequencies $\pm\Omega_p$ in the resonant driving case. Increasing $\Delta_p$ from zero, results in an asymmetric spectrum where amplification occurs only on one sideband while an absorption feature occurs on the other, as shown in Fig. \ref{fig: MollowRb2}. 

Figure\, \ref{fig: Keldysh3sc} shows  various spectral lines corresponding to features in the absorption spectrum for the slightly detuned case  as in Fig. \ref{fig: MollowRb2}. In the atom-photon number dressed state basis, we see that the features correspond to transitions between different dressed states involving absorption/emission of a probe photon with energy $\hbar\omega_{pr}$ and absorption/emission of a number of pump photons with energy $\hbar\omega_p$.

\begin{figure}[h!]\centering
    \includegraphics[scale=0.25]{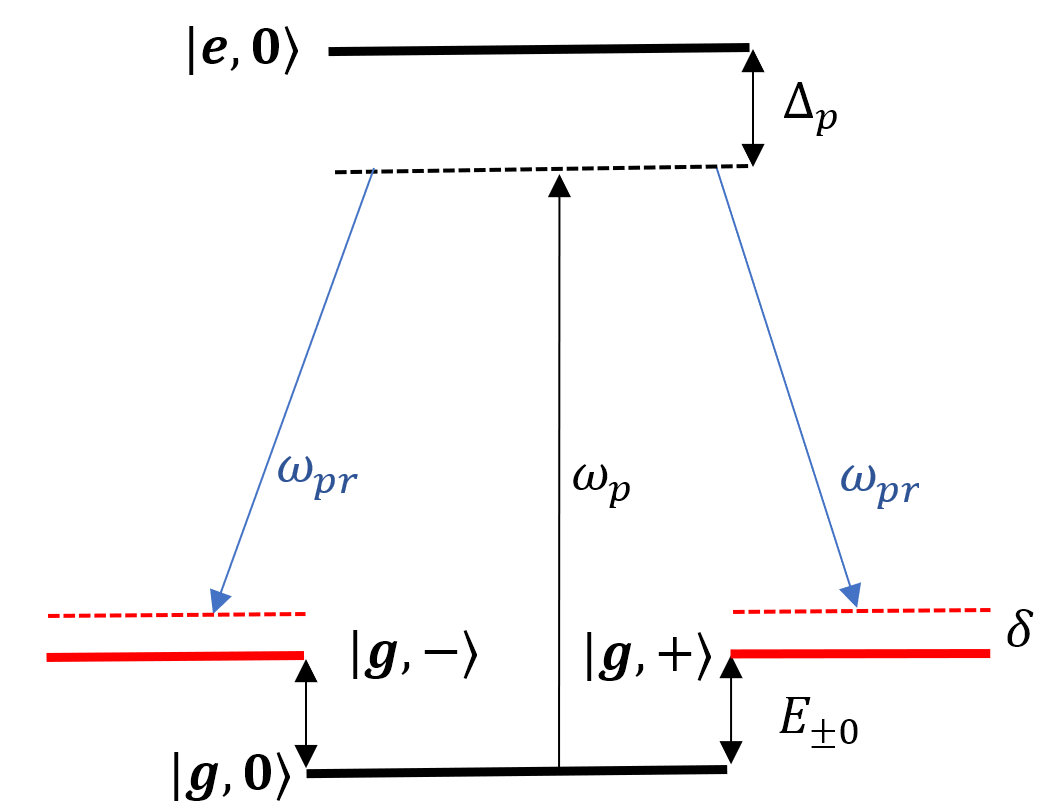}
    \caption{Raman gain in a detuned system with multiple magnetic sub-levels. $\ket{g,m}$ is the ground state magnetic sub-level $m$. The pump field ($\omega_{p}$) and spontaneous decay create a population distribution among the $\ket{g,m}$ states that depends on their Clebsch-Gordon coefficients. The probe laser ($\omega_{pr}$) experiences gain/loss depending on the signs of $\Delta_p$, $\delta$.}
    \label{fig: RamanZ}
\end{figure}

\subsection{Raman gain using Zeeman  sublevels}

\begin{figure}[h]\centering
    \includegraphics[scale=0.25]{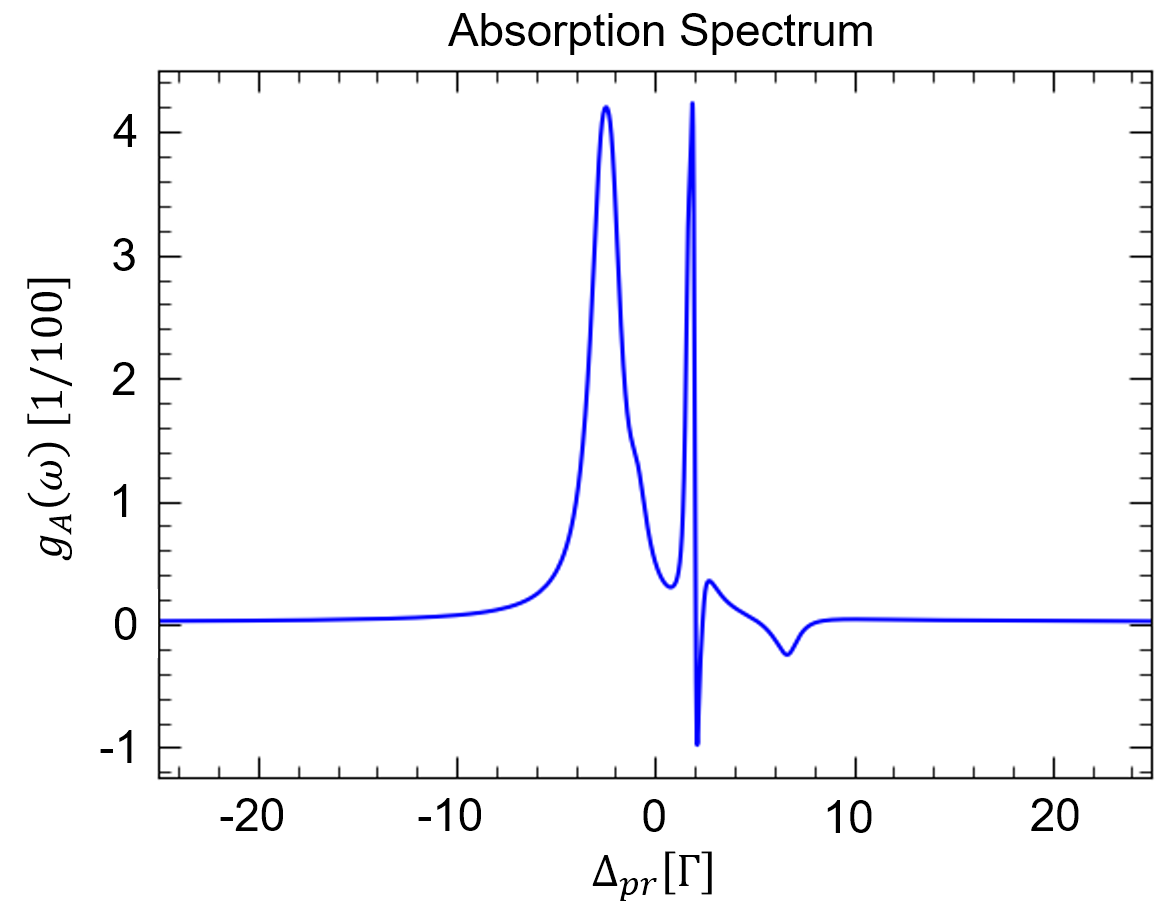}
    \caption{Absorption spectrum of light with perpendicular polarization for the off-resonantly driven $^{87}$Rb system with parameters $\Gamma=1.0$, $\Omega_{p}=4\Gamma$, $\Delta_p=1.75\Gamma$.}
    \label{fig: MollowRb3}
\end{figure}

For the case of a two-level system with many magnetic sub-levels, a far-detuned pump laser field can induce probe gain through population inversion among the  shifted sub-levels. Unlike Mollow gain, Raman gain on magnetic sublevels requires a probe field with polarization orthogonal to the pump filed polarization \cite{Guerin2008}. This is due to the necessary two-photon Raman transition between different $m$ number states.
The high one-photon detuning creates level shifts of the dressed states corresponding to each magnetic sub-level and introduces a dispersive structure for the absorption spectrum at $\delta=0$ shown in Fig.\,\ref{fig: RamanZ}. This results in gain on one side of the structure and loss on the other, shown in Fig.\,\ref{fig: MollowRb3}.

\subsection{Raman gain with coupling fields}
\begin{figure}[h!]\centering
    \includegraphics[scale=0.35]{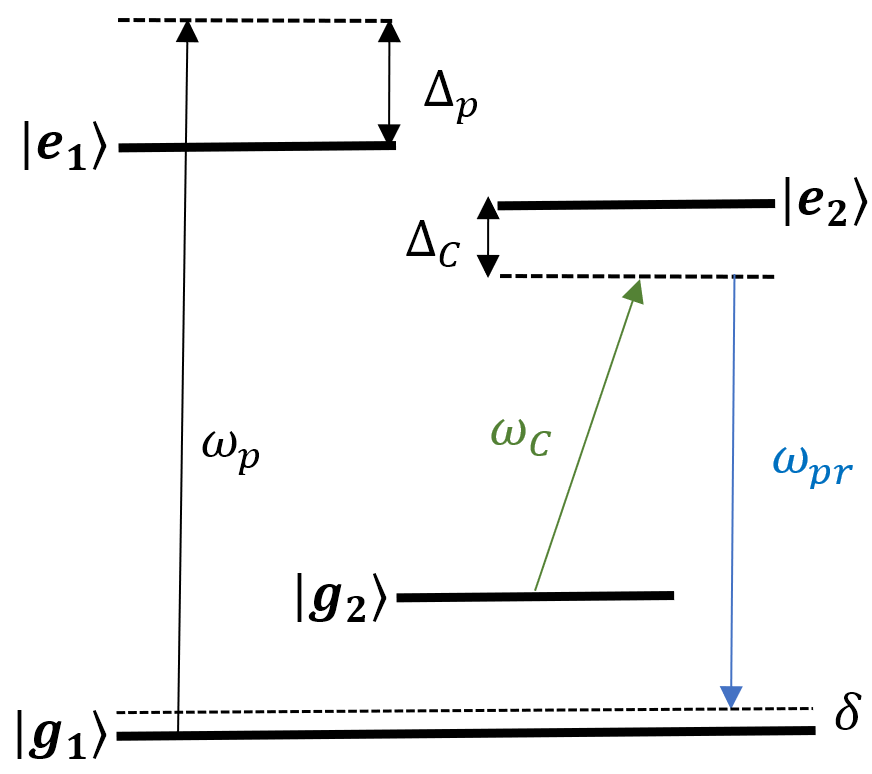}
    \caption{Raman gain in a system with coupled hyperfine levels $g_i$. The pump laser ($\omega_{pr}$) optically pumps the system to induce ground state coherence. The coupling laser ($\omega_C$) is introduced to return atoms to the ground states and create gain in the probe ($\omega_p$) due to the Raman transition involving the coupling and probe.}\label{RamanH}
\end{figure}
Raman gain can be induced by sustaining a population inversion between two different lower energy states, e.g. hyperfine ground states \cite{Baudouin} through optical pumping, and then producing gain with a stimulated two-photon transition using a coupling field $E_c(t)$ \cite{Baudouin2013,Baudouin}. The probe polarization in this case depends on the choice of level $\ket{e_2}$, see Fig.\,\ref{RamanH}.

\subsection{Four-wave mixing (FWM) gain}
FWM is a phase-sensitive parametric nonlinear optical process that involves the interactions of four fields with a nonlinear medium and transferring energy between the fields \cite{Li2012}. When considering a degenerate FWM (d-FWM) using counter-propagating pump fields and satisfying a phase matching condition, gain is observed in the backward propagating reflection of the probe field \cite{Baudouin, Boyd1981}. 

\subsection{Inversionless gain in coherently driven multilevel systems}
Probe gain in coherently driven multilevel systems without inversion in any meaningful basis has been described as resulting from coherence between dressed states \cite{Mompart2000, Agarwal1991}. For the case of the three-level $\text{V}$-system, each of the coherent, linked dressed state contributes to the absorption spectrum with two independent Lorentzians occurring at the dressed states resonances. 
The presence of gain is suggested to be due to competition between n-photon scattering processes, each of which is  responsible for contributing to gain or loss. In $\text{V}$ and $\Lambda$ type systems, two-photon scattering is responsible for inversionless gain,  while one-photon processes cause gain in cascade type systems \cite{Mompart2000}.

\subsection{Amplified Spontaneous Emission (ASE)}
Spontaneous emission is a quantum process in which an atom in an excited state spontaneously de-excites into a lower energy state by emitting a photon into one of the modes of the quantum mechanical vacuum/reservoir field. It can be semiclassically described  as emission process stimulated by vacuum noise, (the zero-point fluctuations of the electric field in the vacuum state) \cite{GeaBanacloche1988}. ASE is a result of stimulated emission processes that amplify a  spontaneous radiation field propagating in a medium with gain and is often a pervasive phenomenon in lasers and optical amplifiers \cite{Lippi2021}. We distinguish between ASE and the cooperative phenomena of superfluorescence, the latter process requiring negligible dephasing to form a cohrent macroscopic interaction \cite{Malcuit87}.

The threshold for ASE in a particular gain medium is reached when the contribution from optical gain processes is greater than that of absorption and escape processes,  (radiation scattering away from the medium). Unlike lasing involving coherent amplification of a probe pulse, the temporal coherence of ASE in media with multiple scatterers without coherent feedback is often low, due to the large radiation bandwidth. In particular, the second-order coherence $g^{(2)}(\omega,t)$ of ASE demonstrates super-Poissonian statistics consistent with that of thermal light \cite{Blazek2011}. However the presence of a coherent feedback loop realizes random lasing and the generation of more coherent light with Poissonian statistics \cite{Doronin2019}. Coherent feedback loops can be created by manipulating the scattering processes in the gain medium, as multiple scattering can replace the requirement of mirrors for coherent feedback.  This leads to the formation of a new threshold for lasing with coherent feedback that depends on the gain medium's scattering properties \cite{Cao2000}.

The spatial coherence of ASE can be very high even without a coherent feedback loop. In atomic gas systems, the geometry of the lasing setup (the optically active part of the system), as well as the internal degrees of freedom (atomic populations, coherences and the spectral lineshape \cite{Hazak1992}), affect spatial coherence and beam divergence. The treatment of spatial coherence becomes more complicated when considering multiple scattering, especially in the strong scattering regime where localization of light can occur \cite{Wiersma2008}.

The starting point for the microscopic treatment of ASE in a continuously driven medium is the Heisenberg equations of motion for the photon number operator Eq. \eqref{Radb}. In general, it is difficult to solve for the photon number operator, especially in the case of large systems. Instead, the operator $\expval{E_{I^{\prime}}^{(\mu)-}(\vec{r})E_{I^{\prime}}^{(\mu^{\prime})+}(\vec{r})}(t)$ is considered. Before we look at the effects of feedback and multiple scattering, we look into this operator in detail. The contribution of  spontaneous emission in $\expval{E_{I^{\prime}}^{(\mu)-}(\vec{r},t)E_{I^{\prime}}^{(\mu^{\prime})+}(\vec{r},t)}$ is
\begin{align}\label{SnExp2}
    \begin{split}
        S_{\text{s}}(t)
        =\sum_{j,\alpha}\sum_{i,\beta;\nu=\pm}\int_0^t dt^{\prime}&
        \left(\Gamma^{(\mu\mu^{\prime})}_{ji}(\vec{R_{\alpha}},\vec{R_{\beta}},t,t^{\prime})\expval{{\sigma}_{i,I^{\prime}}^{+}(\vec{R}_{\alpha},t){\sigma}_{j,I^{\prime}}^{-}(\vec{R}_{\beta},t^{\prime})}\right.\\        &\left.+\Gamma^{(\mu\mu^{\prime})}_{ij}(\vec{R_{\alpha}},\vec{R_{\beta}},t,t^{\prime})\expval{{\sigma}_{j,I^{\prime}}^{+}(\vec{R}_{\beta},t^{\prime}){\sigma}_{i,I^{\prime}}^{-}(\vec{R}_{\alpha},t)}\right),
    \end{split}
\end{align}
where $\Gamma^{(\mu\mu^{\prime})}_{ij}(\vec{R_{\alpha}},\vec{R_{\beta}},t,t^{\prime})$ is given by
\begin{equation}
    \int\int d^3k\text{ }d^3k^{\prime}\text{ }g_{\vec{k};i}^{(\mu)}(\vec{R_{\alpha}},t)g_{\vec{k^{\prime}};j}^{(\mu^{\prime})*}(\vec{R_{\beta}},t^{\prime}).
\end{equation}

We distinguish between the single-atom term ($\alpha=\beta$) and the two-atom terms ($\alpha\not=\beta$). The latter contributes to collective emission, the superfluorescence. 
The stimulated emission term is
\begin{align}\label{SnExp}
    \begin{split}
        \mathcal{R}_{\text{s}}(t)
        =&~i\sum_{i,\alpha}\int\int d^3k\text{ }d^3k^{\prime}\left(g_{\vec{k}^{'}i}^{\left(\mu^{\prime}\right)*}(\vec{R}_{\alpha})\Tr{a_{\vec{k},I^{\prime}}^{(\mu)\dag}(t){\sigma}_{i,I^{\prime}}^{-}(\vec{R}_{\alpha},t)\rho_{I^{\prime}}(t_0)}\right.\\
        &\qquad\left.-g_{\vec{k}i}^{\left(\mu\right)}(\vec{R}_{\alpha})\Tr{a_{\vec{k}^{\prime},I^{\prime}}^{(\mu^{\prime})}(t){\sigma}_{i,I^{\prime}}^{+}(\vec{R}_{\alpha},t)\rho_{I^{\prime}}(t_0)}\right)\\
        &-\sum_{ij,\alpha;\nu=\pm}\int_0^t dt^{\prime}
        \left(\expval{E_{I^{\prime}}^{(\mu)-}(\vec{r},t)\left[\vec{d}_l\cdot \vec{E}_{I^{\prime}}^{(\nu)+}(\vec{R}_{\alpha},t^{\prime})\right]\cdot\mathcal{C}^{(\mu^{\prime})}_{i,j}(\vec{r},\vec{R}_{\alpha},t,t^{\prime})}\right.\\        &\qquad\left.+\expval{\left[\vec{d}^{*}_l\cdot\vec{E}_{I^{\prime}}^{(\nu)-}(\vec{R}_{\alpha},t^{\prime})\right]E_{I^{\prime}}^{(\mu^{\prime})+}(\vec{r},t)\cdot \mathcal{C}^{(\mu)\dag}_{i,j}(\vec{r},\vec{R}_{\alpha},t,t^{\prime})}\right).
    \end{split}
\end{align}
The first term, the Langevin force, is non-zero only if there are initial correlations at time $t_0$. The $\vec{d}_l$ is the unit transition dipole vector for transition $l$. The term proportional to the single-atom dipole correlation operator reads
\begin{align}\label{Dipcor}
    \mathcal{C}_{i,k}^{\mu^{\prime}}(\vec{r},\vec{R}_{\alpha},t,t^{\prime})&=\comm{{\sigma}_{i,I^{\prime}}^{-}(\vec{R}_{\alpha},t)}{{\sigma}_{j,I^{\prime}}^{+}(\vec{R}_{\alpha},t^{\prime})} \times \int d^3k^{\prime}\text{ }g_{\vec{k}^{'}i}^{\left(\mu^{\prime}\right)*}(\vec{R}_{\alpha}-\vec{r})\varepsilon^{(\mu^{\prime})}_{\vec{k^{\prime}}}.
\end{align}
The single-atom dipole-dipole functions are expanded in terms of projection operators onto the populations and coherences to give us the contributions to the spontaneous emission and gain/loss spectra.  In the case where we have non-degenerate levels  and no driving fields, one transition is coupled to each radiation mode, and  single-atom dipole-dipole functions for transition $l$ are equal to sums of $\sigma_l^z$ and  $I_l$. Coupling multiple transitions to a single mode introduces lower-state or upper-state coherence. The contribution from each transition is weighted by their Clebsch-Gordan coefficient and the transition frequency.

In the interaction picture where the atomic propagators are dressed with the driving field, non-zero driving introduces the $\sigma_l^{\pm}$ operators in the dipole-dipole functions, thus introducing dependence on the atomic coherences. In this way, the dressed state structure of the degenerate system appears with the presence of new Lorentzians and interference terms in the spectral function. Since the atomic coupling to the vacuum is weak, it is useful to switch to the more physically meaningful basis of the dressed states  which are the eigenstates of $U_L(t,t_0)$.

We assume that driving fields are slowly varying, with small losses in the intensities over large timescales.
The dressed states, $\ket{\Lambda_k(\vec{R}_{\alpha},t)}$, are defined along with their corresponding energies $\lambda_k(t)$ such that
\begin{equation}\label{Dresexp}
\begin{split}
    &\sigma_{j;I^{\prime}}^{+}(\vec{R}_{\alpha},t)=\sum_{m^{\prime}m}C^{j}_{m^{\prime}m}(\vec{R}_{\alpha},t)e^{it(\omega_j+\lambda_{m^{\prime}}-\lambda_{m})}\ket{\Lambda_m(\vec{R}_{\alpha},t)}\bra{\Lambda_{m^{\prime}}(\vec{R}_{\alpha},t)}.
\end{split}
\end{equation}

The dipole-dipole function consists of projection operators onto the dressed state populations and coherences. 
This causes all fast time dependencies  to be stored in the exponentials and the density matrix. Due to  small coupling strength, and with the assumption that transition frequencies are in the optical regime, the timescale for the evolution of the density matrix is significantly larger  than that for the decay of the correlation functions. This justifies the use of the Markov approximation. 
\subsection{Paraxial approximation}
A pencil-like geometry for the pump laser is assumed such that only a narrow cone of wavevectors $\Sigma_k$, for forwards ($R$) and backwards ($L$)  components, contributes to the ASE modes. The electric field operator for a single mode $\hat{E}^{(\mu)}(\vec{r},t)$ is therefore expanded as:
\begin{equation}\label{Eparax}
    \hat{E}^{(\mu)}(\vec{r},t)=\left[E^{(\mu)+}_R(\vec{r},t)e^{i\vec{k}_0\cdot\vec{r}}+E^{(\mu)+}_L(\vec{r},t)e^{-i\vec{k}_0\cdot\vec{r}}\right]e^{-i\omega_0t}-\text{H.c.}\,,
\end{equation}
where $\vec{k}_0$ is the forward propagating wavevector along the axis of $\Sigma_k$ and $E^{(\mu)+}_{\hat{k}}(\vec{r},t)$ is a slowly varying operator.
This treatment is similar to the description of ASE for the case of homogeneously broadened three-level atoms in a rod-like geometry \cite{Garrison1988}. 
We assume that the atomic density is large enough that the dipole operator is now a continuous function of position. Furthermore we assume in a small volume $V_r$ with radius $r$ centered at any point $\vec{R}$, the $\omega_j c^{-1}r\gg 1$, and the dipole correlation function and the atomic density varies trivially. We introduce  the volume-averaged dipole in a small vertical slice of the cylindrical medium $S_{j;I^{\prime}}^{+}(\vec{R}_{i},t)=n(\vec{R}_{i})\int_{V_r(\vec{R}_i)}dV^{'} \sigma_{j;I^{\prime}}^{+}(\vec{R}^{'},t)$. 
The interaction terms between pairs of atoms are assumed to contribute negligently to the dynamics and are ignored. The equation for the density matrix element in the dressed state basis reads
\begin{equation}
\begin{split}
    \dfrac{d}{dt}\expval{\sigma_{\Lambda_a\Lambda_b}(\vec{R}_a,t,t)}_{I^{'}}
    &=\sum_{ml}\Gamma_{1}^{ab;ml}(\vec{R}_{\alpha},t)\expval{\sigma_{\Lambda_m\Lambda_l}(\vec{R}_a,t,t)}_{I^{'}}\\
    &\quad-\sum_{l}\Gamma_{2}^{ab;l}(\vec{R}_{\alpha},t)\expval{\sigma_{\Lambda_a\Lambda_l}(\vec{R}_a,t,t)}_{I^{'}}\\
    &\quad-\sum_{l}\Gamma_{3}^{ab;l}(\vec{R}_{\alpha},t)\expval{\sigma_{\Lambda_l\Lambda_b}(\vec{R}_a,t,t)}_{I^{'}}\\
    &\quad+\sum_{\hat{k}_1,\hat{k}_2}I^{(\mu\mu^{\prime})}(\hat{k}_1,\hat{k}_2,\vec{R}_{\alpha},\vec{R}_{\alpha},t)\\
    &\qquad\times\left(\sum_{ml}B_{1}^{ab;ml}(\mu,\mu^{\prime},\vec{R}_{\alpha},t)\expval{\sigma_{\Lambda_m\Lambda_l}(\vec{R}_a,t,t)}_{I^{'}}\right.\\
    &\qquad\left.+\sum_{ml}B_{2}^{ab;l}(\mu,\mu^{\prime},\vec{R}_{\alpha},t)\expval{\sigma_{\Lambda_a\Lambda_l}(\vec{R}_a,t,t)}_{I^{'}}\right.\\
    &\qquad\left.+\sum_{ml}B_{3}^{ab;l}(\mu,\mu^{\prime},\vec{R}_{\alpha},t)\expval{\sigma_{\Lambda_l\Lambda_b}(\vec{R}_a,t,t)}_{I^{'}}\right).
\end{split}
\end{equation}

We define the two-point field correlation function as $I^{(\mu\mu^{\prime})}(\hat{k}_1,\hat{k}_2,\vec{R}_{1},\vec{R}_{2},t)=\expval{E_{\hat{k}_1,I^{\prime}}^{(\mu)-}(\vec{R}_{1},t)\vec{E}_{\hat{k}_2,I^{\prime}}^{(\mu^{\prime})+}(\vec{R}_{2},t^{\prime})}$. The term $\vec{R}_{1,2}$ is the average polarization vector of the $\mp$ field component and $\hat{k}_{1,2}\in\{L,R\}$. We use the paraxial approximation formalism described in Ref.\,\cite{Garrison1988} to derive the radiative transport equation
\begin{equation}\label{ParRadf}
\begin{split}
    \left[\dfrac{\partial}{\partial t}+c\left(\hat{k}_{1}\cdot\nabla_{\vec{R}_1}+\hat{k}_{2}\cdot\nabla_{\vec{R}_2}\right)\right.\qquad\quad&\\
    \left.-i\dfrac{c}{2k_0}\nabla_{\perp}^2\right]I^{(\mu\mu^{\prime})}(\hat{k}_1,\hat{k}_2,\vec{R}_{1},\vec{R}_{2},t)&=\sum_{\vec{x},ij}A^{ij;ml}_{\mu\mu^{\prime}}(\vec{x},t)\expval{S_{\Lambda_m\Lambda_l}(\vec{x},t,t)}_{I^{'}}\\
    &~~+\sum_{\hat{q}\in\{L,R\}}\sum_{\vec{x},ij}\delta_{\mu\mu^{\prime}}\expval{S_{\Lambda_m\Lambda_l}(\vec{x},t,t)}_{I^{'}}\\
    &~~\times \left[K^{ij;ml}(\hat{q}\rightarrow\hat{k}_1,\hat{k}_2,\vec{x},t)I^{(\mu\mu)}(\hat{q},\hat{k}_2,\vec{R}_1,\vec{R}_2,t)\right.\\
    &~~\left.+K^{ij;ml}(\hat{k}_1,\hat{q}\rightarrow\hat{k}_2,\vec{x},t)I^{(\mu\mu)}(\hat{k}_1,\hat{q},\vec{R}_1,\vec{R}_2,t)\right],
\end{split}
\end{equation}
where $\nabla_{\perp}$ is the gradient operator transverse to $\hat{k}_{1,2}$ and $A^{ij;ml}_{\mu\mu^{\prime}}(\vec{x},t)$ is given by
\begin{equation}
\begin{split}
    A^{ij;ml}_{\mu\mu^{\prime}}(\vec{x},t)&= \int_{\Sigma_k}\int_{\Sigma_{k^{\prime}}} d^3k\text{ }d^3k^{\prime}\text{ }g_{\vec{k};i}^{(\mu)*}(\vec{x})g_{\vec{k^{\prime}};j}^{(\mu^{\prime})}(\vec{x})\\
    &~~\times\sum_{l^{\prime}}\int_{0}^{\infty} d\tau\bigg[C_{l^{\prime}l}^{i*}(t)C_{ml^{\prime}}^{j}(t)e^{-i(\omega_k-\omega_{k^{\prime}}-\omega_{ij}+\lambda_{l^{\prime}l}-\lambda_{l^{\prime}m})t}~e^{i(\omega_{k^{\prime}}-\omega_{j}+\lambda_{l^{\prime}m})\tau}\\
    &\qquad\qquad\qquad\qquad+C_{ml^{\prime}}^{i}(t)C_{l^{\prime}l}^{j*}(t)e^{i(\omega_k-\omega_{k^{\prime}}-\omega_{ij}-\lambda_{l^{\prime}l}+\lambda_{l^{\prime}m})t}~e^{-i(\omega_{k^{\prime}}-\omega_{j}-\lambda_{l^{\prime}m})\tau}\bigg].
\end{split}
\end{equation}

The spontaneous emission contribution is integrated over the cone in momentum space that contains wave-vectors that remain in the cylindrical medium and contribute to ASE. The remaining wavevectors are considered to have escaped from the medium and do not contribute to the evolution of the paraxial ASE modes. $K^{ij;ml}(\hat{q}\rightarrow\hat{k}_1,\hat{k}_2,\vec{x},t)$ and $K^{ij;ml}(\hat{k}_1,\hat{q}\rightarrow\hat{k}_2,\vec{x},t)$ are the scattering probabilities for the mode with wavevector $\hat{q}$ to scatter into the mode with wavevector $\hat{k}_{1,2}$ respectively through virtual transitions between dressed states $\Lambda_m$ and $\Lambda_l$. This corresponds to stimulated emission processes and is derived from Eq.\,\eqref{SnExp} using the dressed state expansion of the atomic operators Eq.\,\eqref{Dresexp}.

The paraxial approximation equations for the dressed state atomic matrix and the radiative equations for the field correlation effectively model the dynamics of an ASE field in a cold random atomic gas, under the approximation that interatomic scattering plays little to no effect. The gain condition that results in ASE can be derived by calculating all the coefficients that depend on the dressed state configuration and determining the choices of parameters, including the pump field and detuning, that result in the right-hand side of Eq.\,\eqref{ParRadf} being greater than zero  in the steady state. The choice of parameters can be informed by looking at the gain mechanisms, and the above  section on the probe field gain  accounts for how dressed state inversions and coherences contribute.

\section{Degenerate Mirrorless Lasing}\label{sec:mirrorlessLasing}

In this section, we study a prototypical system in which directional emission was predicted and observed to occur at the same frequency and orthogonal polarization as those of the excitation light.

Consider the eight-level system of the $D_2$ ($5^2S_{1/2} \rightarrow 5^2P_{3/2}$) line in $^{87}$Rb. 
The $5^2S_{1/2}$ and $5^2P_{3/2}$ states are split into hyperfine structure components with total angular momentum $F_g=2,1$ and $F_e=3,2,1,0$ respectively. We consider the $F_g=1 \rightarrow F_e=2$ transition. In the absence of magnetic field, the ground state is three-fold  and the excited state is five-fold degenerate.

\begin{figure}[h!]
	\centering	\includegraphics[scale=0.6]{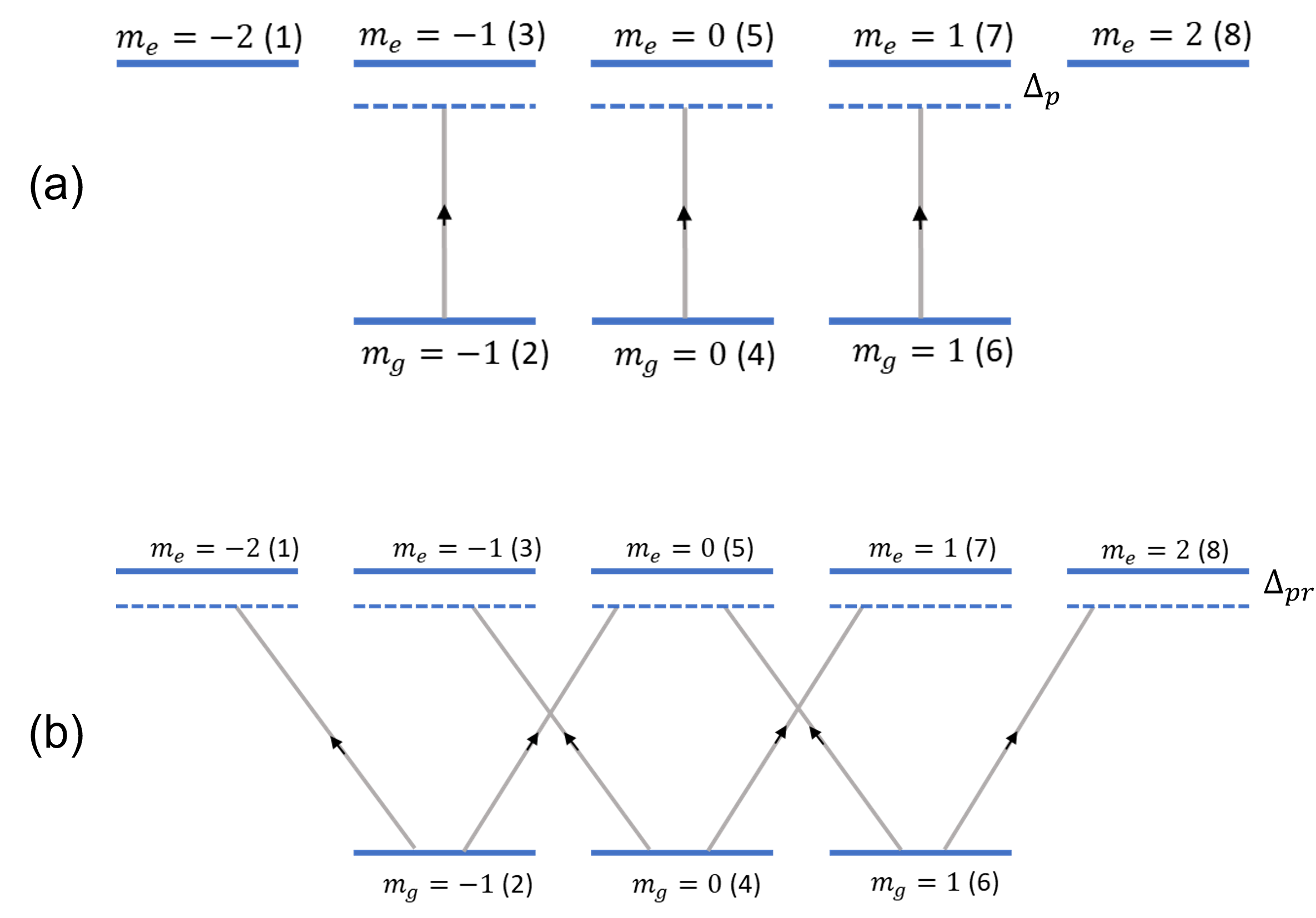}
	\caption{The degenerate 8-level system interacting with a linearly polarized pump field in $z$-direction (a) and linearly polarized probe field in the orthogonal direction (b). The numbers given in parenthesis are the numbers corresponding to the levels as they appear in the Hamiltonian.} \label{8-level}
\end{figure}

Pump light linearly polarized in the $z$-direction induces the transition between magnetic sublevels with the same quantum numbers: $m_g=m_e$. On the other hand, circularly polarized light induces transition between energy level with different magnetic quantum numbers depending on the direction of polarization. This is due to the principle of angular momentum conservation, as the sum of angular momenta of the light and the atom should remain constant. If $E_{\sigma_-}$ and $E_{\sigma_+}$ represent the right and left circular polarization components of the light, a superposition of the two components chosen here as the probe gives linearly polarized field in the $\hat{x}$ direction
\begin{eqnarray}
	E_x = \frac{i}{\sqrt{2}} (E_{\sigma_-}+E_{\sigma_+})\, ,
\end{eqnarray}
where the overall phase factor is chosen for consistency with standard definitions.

\subsection{Population Dynamics}
The interaction of the mutually orthogonal linearly polarized pump and probe fields with atoms is depicted in Fig.\,\ref{8-level}. The pump field polarized along the quantization axis $z$ couples the ground and excited states satisfying the condition $\Delta m= m_e-m_g =0$, as shown in Fig.\,\ref{8-level}(a). The orthogonally polarized probe field couples states satisfying $\Delta m= m_e-m_g =\pm 1$, see \ref{8-level}(b).

The field-interaction Hamiltonian is given by:
\begin{widetext}
	\begin{equation}\label{HAM_linear}
		H = \frac{\hbar}{2}
		\begin{pmatrix}
			2\Delta_{pr}	&	-i \Omega_{21(pr)}	&	0	&	0	&	0	&	0	&	0	&	0\\
			i \Omega_{21(pr)}	&	0	&	\Omega_{23(p)}	&	0	&	i \Omega_{25(pr)}	&	0	&	0	&	0\\
			0	&	\Omega_{23(p)}	&	2(\Delta_p+\Delta_{pr})	&	-i \Omega_{43(pr)}	&	0	&	0	&	i \Omega_{47(pr)}	&	0\\
			0	&	-i \Omega_{25(pr)}	&	0	&	0	&	\Omega_{45(p)}	&	0	&	i \Omega_{47(pr)}	&	0\\
			0	&	-i \Omega_{25(pr)}	&	0	&	\Omega_{45(p)}	&	2(\Delta_p+\Delta_{pr})	&	-i \Omega_{65(pr)}	&	0	&	0\\
			0	&	0	&	0	&	0	&	i \Omega_{65}	&	0	&	\Omega_{67(p)}	&	i \Omega_{68(pr)}\\
			0	&	0	&	0	&	-i \Omega_{47}	&	0	&	\Omega_{67(p)}	&	2(\Delta_p+\Delta_{pr})	&	0\\
			0	&	0	&	0	&	0	&	0	&	-i \Omega_{68(pr)}	&	0	&	2\Delta_{pr}\\
		\end{pmatrix}\,,
	\end{equation}
\end{widetext}
where $\Delta_p=\omega_{0}-\omega_p$ and $\Delta_{pr}=\omega_{0}-\omega_{pr}$ are pump and probe detunings, respectively, and $\Omega_{ij(p,pr)}$ are the Rabi frequencies for transitions between $i$ and $j$ sublevels corresponding to either the pump or the probe field. The Rabi frequencies are given by
\begin{eqnarray}
	\Omega_{ij(p,pr)}	&=& -\frac{E_{p,pr} \mel{F_g\,m_g}{e\mathbf{r}}{F_e\,m_e}}{\hbar}\,,
\end{eqnarray}
with the interaction strengths characterized by the dipole matrix elements between a ground state $\ket{F_g\,m_g}$ and an excited state $\ket{F_e\,m_e}$. To simplify the calculation of Rabi frequencies, the matrix elements are expressed in terms of the 3-j sysmbol so that the angular momentum dependence can be factored out:
\begin{eqnarray}\label{cleb_3j}
	\mel{F_g\,m_g}{e\mathbf{r}}{F_e\,m_e} = \bra{F_g}|e\mathbf{r}|\ket{F_e}(-1)^{F_e-1+m_g} \times \sqrt{2F_g+1}\begin{pmatrix}
		F_e & 1 & F_g\\
		m_e & q & -m_g
	\end{pmatrix}\,,
\end{eqnarray}
where $\bra{F_g}|e\mathbf{r}|\ket{F_e}$ is the reduced matrix element, independent of the magnetic quantum numbers
and $q$ is the index of $\mathbf{r}$ in the spherical basis. Note that the reduced matrix element will be the same for all transitions. 
The pump (probe) reduced Rabi frequency $\Omega_p$ ($\Omega_{pr}$) can now be defined in terms of the reduced matrix element, such that
\begin{eqnarray}
	\Omega_{p,pr}	&=& -\frac{E_{p,pr} \bra{F_g}|e\mathbf{r}|\ket{F_e}}{\hbar}.
\end{eqnarray}

Any population in the excited state decays to the ground sublevels satisfying the selection rules, $\Delta m = 0, \pm 1$. The decay is characterized by the branching ratio. Considering that the total decay rate of the excited state is $\Gamma$, the density matrix equation is given by
\begin{equation}\label{master_equation}
	\frac{d}{dt}\boldsymbol{\rho} = -\frac{i}{\hbar} [\mathbf{H},\boldsymbol{\rho}]-\frac{\Gamma}{2}\sum_{k=1}^{3} \left(\boldsymbol{\sigma}_k^+\boldsymbol{\sigma}_k^-\boldsymbol{\rho} + \boldsymbol{\rho} \boldsymbol{\sigma}_k^+\boldsymbol{\sigma}_k^- -2\boldsymbol{\sigma}_k^- \boldsymbol{\rho} \boldsymbol{\sigma}_k^+\right) \,,
\end{equation}
where $\boldsymbol{\sigma}^+_k$ and $\boldsymbol{\sigma}_k^-$ are the raising and lowering operators for decay channels with $\Delta m=0$, $\Delta m = -1$ and $\Delta m =+1$  denoted by $k=1,2,3$, respectively. 

\begin{figure}
	\centering    \includegraphics[scale=0.5]{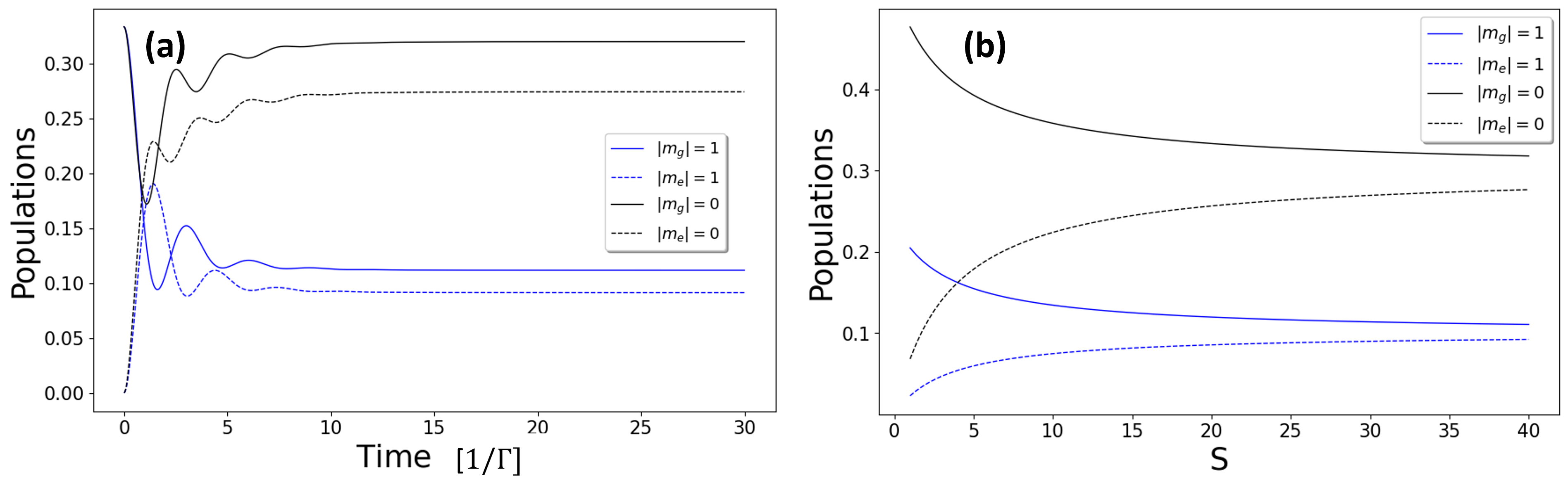}
	\caption{The evolution of populations for $S=36$, (a), and the steady-state populations vs saturation parameters (b) in the presence of linearly polarized pump field. Under the combined effect of strong pumping and decay, the populations tend to move towards the hyperfine levels with lower magnetic numbers. While there is no inversion of population between the $m_g = 0$ and $m_e = 0$ levels, there is a population inversion between $|m_g|=1$ and $m_e=0$ levels. Since the pump is linearly polarized, the $|m_e|=2$ levels are not populated at all. Here, $\Delta_p=0$ in both (a) and (b).} \label{population_inversion}
\end{figure}

First, consider that the population is initially distributed equally in the ground sublevels and a cw pump field, linearly polarized in the $z$-direction, is applied. It induces population transfer with $ \Delta m = 0$. Consequently, the states with $|m_e| = 2$ do not obtain any population. The combined process of strong pumping and decay moves populations to sublevels with lower magnetic numbers and eventually brings the system into a steady-state. A useful quantity called saturation parameter, denoted by S, can be defined in terms of the reduced Rabi frequency, spontaneous decay rate and detuning
\begin{equation}
	S = \frac{\Omega_p^2}{\frac{\Gamma^2}{4}+\Delta_p^2}\,.
\end{equation}
As mentioned earlier, the frequency parameters are expressed in units of $\Gamma$ and time parameters in the units of $1/\Gamma$.
An example of the time evolution of the populations for $S=36$ is shown in Fig.\,\ref{population_inversion}(a). Initially,  populations are equally distributed among the three ground degenerate levels. As the system reaches the steady-state, a population inversion between $|m_e| = 0$ and $|m_g| = 1$ levels has been achieved.
The one-photon resonance, $\Delta_p=0$, is favorable for achieving population inversion at lower peak Rabi frequency.
In Fig.\,\ref{population_inversion}(b), the steady-state populations vs. saturation parameter are depicted. The population inversion is achieved for values of $S>4.0$, implying a  strong pumping is needed for inversion to occur. The pump field is taken here to be in resonance with the transition in both figures \ref{population_inversion}(a) and \ref{population_inversion}(b). 
Taichenachev et al. \cite{Taichenachev_1996} have derived equations for the steady state populations in a system with $j_g=j$ to $j_e = j+1$ energy levels, coupled resonantly to a pump field. Those steady state equations were derived under the condition that the time derivatives of the density matrix elements are zero. They were used in Ref.\,\cite{movsisyan} to calculate populations vs. saturation parameter for the $F_g = 2 \rightarrow F_e = 3$ and $F_g = 3\rightarrow F_e = 4$ transitions in $^{87}$Rb and $^{85}$Rb respectively. Figure\,\ref{population_inversion} shows the populations obtained from the exact solution of the Liouville von Neuman equation with relaxation; they are in a good agreement with the ones obtained using steady state equations from \cite{Taichenachev_1996} for the $F_g = 1\rightarrow F_e = 2$ case.

\subsection{The probe field gain}
Our objective is to find the conditions for positive gain for probe light polarized orthogonally to the pump field. To achieve this, an equation for amplification (or absorption) of the field is derived from the density matrix equation.  Every atom that is de-excited from the upper state emits a photon, which means
\begin{eqnarray} \label{intensity_photon0}	-\frac{d}{dt}n(\rho^{ex})=\frac{d}{dt}n_p\,,
\end{eqnarray}
where $\rho^{ex}$ stands for the population of the excited state manifold $\rho^{ex}=\sum_{i=1,3,5,7,8}\rho_{ii}$\,; $n$ and $n_p$ are the atom and photon densities, respectively.
Only the photons that are spontaneously emitted within the solid angle $\Phi$ will contribute to the intensity in both forward and backward directions. Considering this, the equation for the photon density is written to account for photons emitted due to stimulated and spontaneous emissions separately
\begin{equation}\label{eq:photons_spont}
	n_p=n_p^{st} + \frac{\Phi}{4\pi}n_p^{sp}\,.
\end{equation} 
The light intensity  is related to the photons density by $I=n_p c\hbar\omega$
where $c$ is the speed of light and $\hbar \omega$ is the energy of a photon. From Eq.\,\eqref{master_equation}, one can identify that the first term is related to stimulated photons while the second term is responsible for spontaneous emissions. Considering this fact and making use of equality $dz = cdt$,
the coupled equations for the propagation of the pump and the probe fields are written as
\begin{equation} \label{propagation_equations}
\begin{split}
\frac{d}{dy}I_z &= -\alpha_zI_z + \frac{\Phi}{4\pi} n\hbar \omega \Gamma_z, \\
\frac{d}{dy} I_x&= -\alpha_xI_x + \frac{\Phi}{4\pi} n\hbar \omega \Gamma_x, 
\end{split}
\end{equation}
where
\begin{equation}\label{alpha111}
\begin{split} 
	\alpha_z &= -\frac{n\omega}{2c\epsilon_0E_{z0}} \sum_{i=1,3,5,7,8} i [\boldsymbol{\mu}_z,\boldsymbol{\rho}]_{ii}  \,,\\
	\alpha_x &= -\frac{n\omega}{2c\epsilon_0E_{x0}} \sum_{i=1,3,5,7,8} i [\boldsymbol{\mu}_x,\boldsymbol{\rho}]_{ii} \,,\\
	\Gamma_z &= \sum_{i=1,3,5,7,8}  \frac{\Gamma}{2}(\boldsymbol{\sigma}_1^+\boldsymbol{\sigma}_1^-\boldsymbol{\rho} + \boldsymbol{\rho} \boldsymbol{\sigma}_1^+\boldsymbol{\sigma}_1^- -2\boldsymbol{\sigma}_1^- \boldsymbol{\rho} \boldsymbol{\sigma}_1^+)_{ii} \,,\\
	\Gamma_x &= \sum_{i=1,3,5,7,8} \sum_{k=2,3}  \frac{\Gamma}{2}(\boldsymbol{\sigma}_k^+\boldsymbol{\sigma}_k^-\boldsymbol{\rho} + \boldsymbol{\rho} \boldsymbol{\sigma}_k^+\boldsymbol{\sigma}_k^- -2\boldsymbol{\sigma}_k^- \boldsymbol{\rho} \boldsymbol{\sigma}_k^+)_{ii} \,.
\end{split}
\end{equation}
Here, $\mu_z$ and $\mu_x$ are the matrices with dipole elements corresponding to the transitions of pump and probe fields respectively.
\begin{equation}
\begin{split}
    	\boldsymbol{\mu}_z &= \mu_{23}\ketbra{2}{3} + \mu_{45}\ketbra{4}{5} + \mu_{67}\ketbra{6}{7} +  \text{H.c.}\,,\\
    \boldsymbol{\mu}_x &= i \Big( \mu_{21}\ketbra{2}{1} + \mu_{25}\ketbra{2}{5} + \mu_{43}\ketbra{4}{3} + \mu_{47}\ketbra{4}{7} + \mu_{65}\ketbra{6}{5} + \mu_{68}\ketbra{6}{8} \Big) + \text{H.c.}
\end{split}
\end{equation}

The spontaneous decay factors $\Gamma_z$ and $\Gamma_x$ are given by
\begin{equation}
\begin{split}
	\Gamma_z &= \Gamma \Big( b_{32}\rho_{33} + b_{54}\rho_{55} + b_{76}\rho_{77} \Big)\,, \\
    \Gamma_x &= \Gamma \Big(b_{12}\rho_{11} + b_{34}\rho_{33} + (b_{52} + b_{56})\rho_{55} + b_{74}\rho_{77} + b_{86}\rho_{88} \Big)\,,
\end{split}
\end{equation}
where $b_{ij}$ are the branching ratios corresponding to the transitions between $i$ and $j$ states. They read
\begin{equation}
\begin{split}
b_{32} &= b_{76} = 1/2 \,, b_{54} = 2/3 \,,\\
b_{12} &= b_{86} = 1\,,\\
b_{34} &= b_{74} = 1/2\,,\\
b_{52} &= b_{56} = 1/6\,.
\end{split}
\end{equation}
\begin{figure}[h!]
	\centering	\includegraphics[width=\columnwidth]{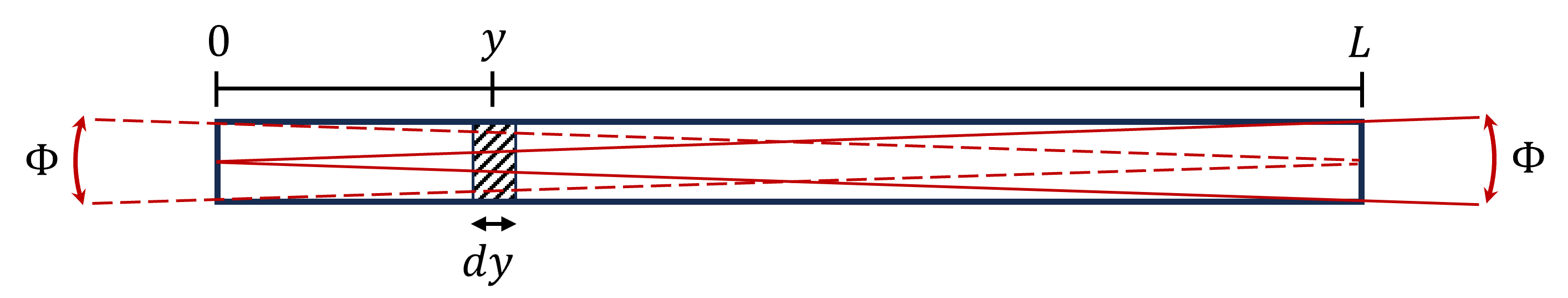}
	\caption{The geometry of the elongated medium responsible for amplified spontaneous emission whose emission solid angle at both ends of the medium is denoted as $\Phi$.} \label{ASE_cell}
\end{figure}
The second term of Eq.\,\eqref{propagation_equations} gives the contribution to the change in intensity due to spontaneous emission of photons of energy $\hbar \omega$ by $n$ atoms per unit volume being in the excited state 
and decaying to respective ground sub-states connected by $z$-polarized or $x$-polarized light. Since the spontaneous decay can be assumed to be isotropic, the factor ${\Phi}/{4\pi}$
accounts for the fraction of spontaneously emitted photons into solid angle $\Phi$.
The solutions of Eqs. \eqref{propagation_equations} are
\begin{equation}
\begin{split}  
	I_z(y) &= I_{z_0} e^{-\alpha_z y} - \frac{\Phi}{4\pi} n\hbar \omega \Gamma_z \left(\frac{e^{-\alpha_z y}-1}{\alpha_z} \right) \,, \\
	I_x(y) &= I_{x_0} e^{-\alpha_x y} - \frac{\Phi}{4\pi} n\hbar \omega \Gamma_x \left(\frac{e^{-\alpha_x y}-1}{\alpha_x} \right)\,,
\end{split}
\end{equation}
where $I_{z_0}=I_z(0)$ and $I_{x_0}=I_x(0)$. In the absence of contribution from spontaneous emission, the amplification (or absorption) of the probe field is determined by the absorption coefficient $\alpha_x$. To analyze this coefficient, it is useful to expand the commutator
\begin{equation}
\begin{split} 
	\sum_{i=1,3,5,7,8} i[\boldsymbol{\mu}_x,\boldsymbol{\rho}]_{ii} &=
	2 \Big( \mu_{21} \Re[\rho_{12}] + \mu_{25} \Re[\rho_{25}] +\mu_{43} \Re[\rho_{34}]\\
	& \qquad+ \mu_{47} \Re[\rho_{47}] + \mu_{65} \Re[\rho_{56}]+\mu_{68}\Re[\rho_{68}] \Big)
\end{split}
\end{equation}
Because of the symmetry in the system, $\rho_{21}=\rho_{68}$,  $\rho_{25}= \rho_{65}$ and $\rho_{43}= \rho_{47}$; the previous equation can hence be reduced to
\begin{eqnarray}\label{eq:commutator}
	\sum_{i=1,3,5,7,8} i[\boldsymbol{\mu}_x,\boldsymbol{\rho}]_{ii} 
	=4 \Big( \mu_{21} \Re[\rho_{12}] +\mu_{43} \Re[\rho_{34}] + \mu_{65} \Re[\rho_{56}] \Big)
\end{eqnarray}

\begin{figure}[h!]
	\centering	\includegraphics[width=\columnwidth]{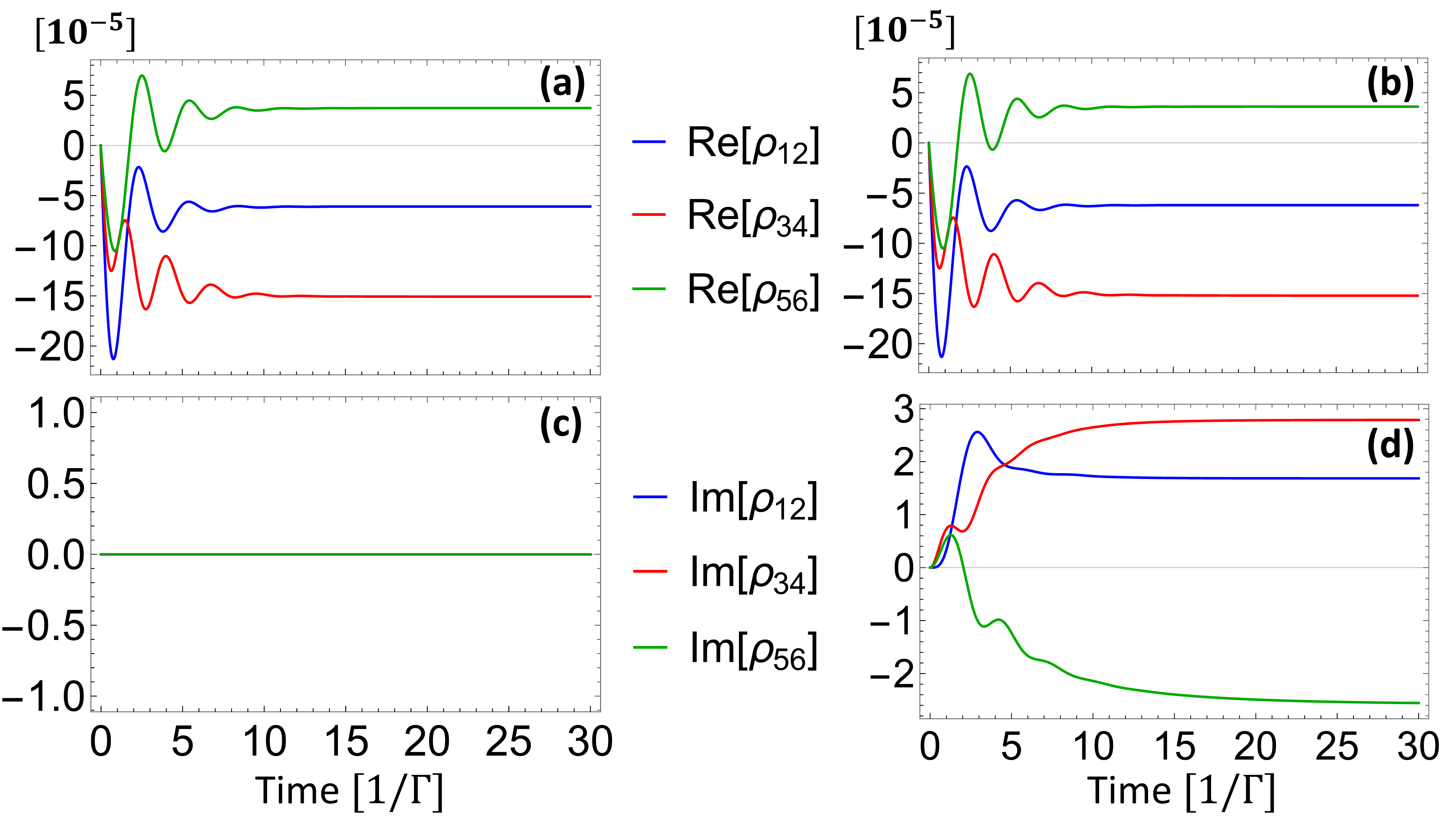}
	\caption{Coherence between states connected by the probe field. In (a) and (c), $\Delta_p=0$, and in (b) and (d), $\Delta_p=0.1 \Gamma$. When the pump field is in resonance, imaginary parts of all the coherences are zero. In the presence of detuning, non-zero imaginary parts contribute to a phase change of the field.} \label{Coherence_probe}
\end{figure}
\begin{figure}[h!]
	\centering
	\includegraphics[scale=0.6]{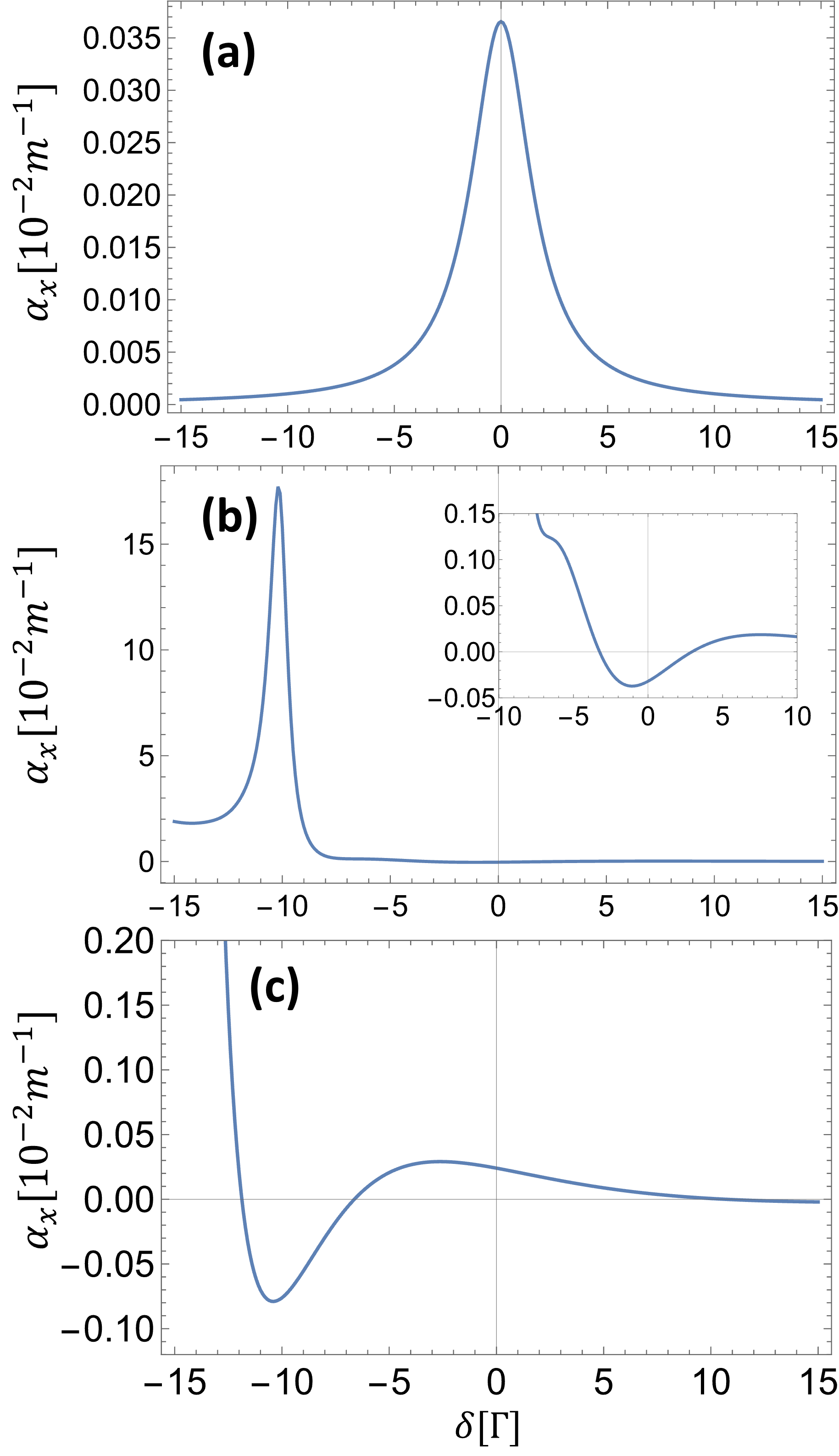}
	\caption{The probe field  absorption as a function of the detuning $\delta$ for different pump detunings considering a very weak probe field, $\Omega_{pr} = 10^{-3}\,\Omega_p$. The pump detunings are (a) $\Delta_p=0$, (b) $\Delta_p=10\,\Gamma$ and (c) $\Delta_p=15\,\Gamma$. A part of the plot in (b) is magnified and shown in the inset. While the absorption is positive for all values in (a), it is negative around 0 in (b) and around $-10\,\Gamma$ in (c), indicating that a high detuning offset is needed for the amplification of the probe field due to stimulated emission. 
    }\label{absoption_spectra}
\end{figure}
The sign of the absorption coefficient $\alpha_x$ in Eq.\,\eqref{alpha111} is opposite to the sign of the quantity 
in Eq.\,\eqref{eq:commutator}. For the probe field to be amplified, $\alpha_x$ needs to be negative. In Fig.\,\ref{Coherence_probe}, the real and imaginary parts of the three coherences $\rho_{12}, \rho_{34}$ and $\rho_{56}$ are plotted for $\Omega_p = 3.0\,\Gamma$ assuming a negligible probe intensity. Figures \ref{Coherence_probe}\,(a) and \ref{Coherence_probe}\,(b) correspond to the resonant case ($\Delta_p = 0$), and \ref{Coherence_probe}\,(c) and \ref{Coherence_probe}\,(d) correspond to the detuned case $\Delta_p=0.1\,\Gamma$. The sum of the real parts is negative in both cases, leading to the coefficient $\alpha_x$ being positive. The positive value of the real part of coherence $\rho_{56}$ between inverted magnetic sublevels reduces the overall negative value of the sum, thus reducing the degree of absorption. Presumably, there exists a system in which the inverted state manifold makes dominant contribution, thus, providing gain in the medium. The present case implies that there is no gain in the medium. Note that all imaginary parts are zero in the resonant case. In the detuned case, imaginary parts have non-zero values. However, they do not contribute to the absorption coefficient; instead they contribute to the change in phase of the electric field.
\begin{figure}[h!]\centering
    \includegraphics[scale=0.25]{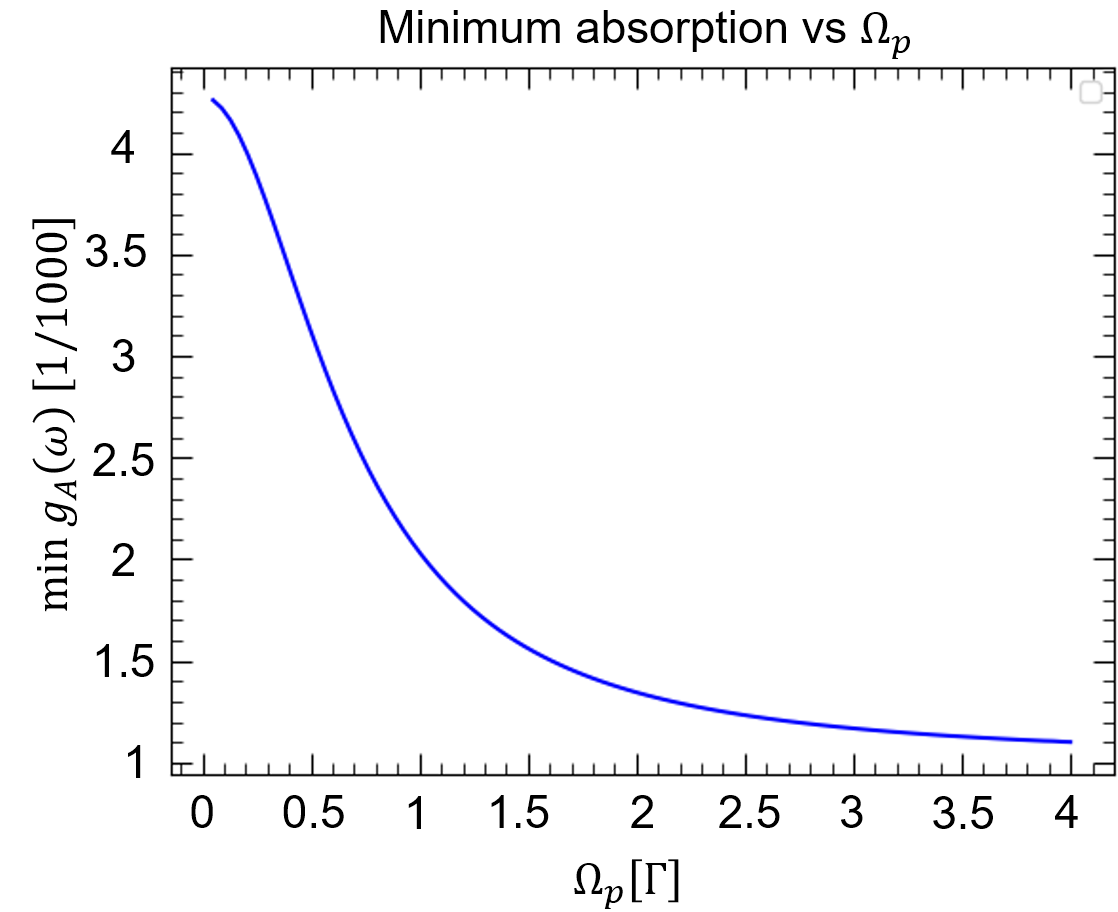}
    \caption{Graph of minimum absorption value over the entire spectrum of radiation for the orthogonally polarized (to the pump) field (seeded by the probe) versus the pump Rabi frequency. The $^{87}$Rb $F=2\rightarrow F=3$ system is used with  parameters $\Gamma=1.0$, $\Delta_p=0$. The probe field frequency that corresponds to the minimum absorption feature varies with $\Delta_p$ and $\Omega_p$. Under the condition of the one-photon resonance there is no gain for any value of $\Omega_p$.}
    \label{fig: Minabsperpres}
\end{figure}

\begin{figure}[h!]\centering
    \includegraphics[scale=0.25]{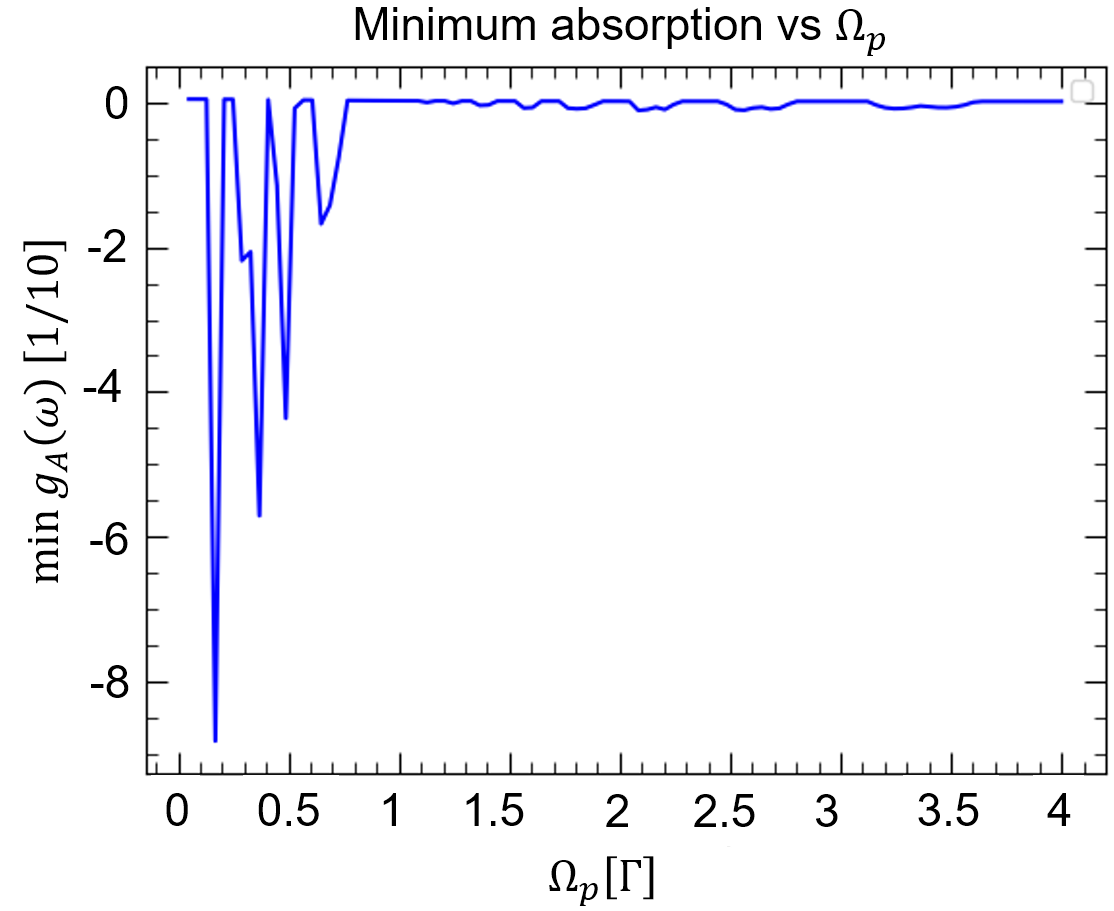}
    \caption{Graph of minimum absorption value over the entire spectrum of radiation for the orthogonally polarized (to the pump) field (seeded by the probe) versus the pump Rabi frequency. The $^{87}$Rb $F=2\rightarrow F=3$ system is used with  parameters $\Gamma=1.0$ and $\Delta_p=0.75\,\Gamma$. Gain is observed when the one-photon detuning is non-zero. }
    \label{fig: Minabsperp075}
\end{figure}

To analyze how detuning impacts the absorption coefficient $\alpha_x$, it is useful to define a quantity called frequency offset given by $\delta = \omega_{p}-\omega_{pr}$. 
The absorption coefficient, considering a very weak probe field $\Omega_{pr} = 10^{-3}\,\Omega_p$, is plotted against $\delta$ for different pump detunings in Fig.\,\ref{absoption_spectra}. The pump detunings are $\Delta_p=0, 10\,\Gamma, 15\,\Gamma$ in Figs.\,\ref{absoption_spectra}\,(a), \ref{absoption_spectra}\,(b) and \ref{absoption_spectra}\,(c) respectively. It is to be noted that there is no gain in the resonant case as well as for small values of detuning offset $\delta$. For higher values of pump detuning there is negative absorption for a certain  range of $\delta$ which is a signature of gain. Such an outcome strongly depends on the value of the Rabi frequency.

The minimum spectral response 
as a function of the Rabi frequency of the pump field is shown in Fig.\,\ref{fig: Minabsperpres}  for the resonant ($\Delta_p=0$) and Fig.\,\ref{fig: Minabsperp075} for the detuned case ($\Delta_p=0.75\,\Gamma$).  They are calculated  using the two-time correlation function in Eq.\,\eqref{spectraltt}. 
The \texttt{Qutip} function library was used to calculate the spectrum using fast Fourier transform of the two-time correlation function. Under resonance, we find that there is no gain for any value of the Rabi frequency. Meanwhile, non-zero detuning provides gain for certain values of the Rabi frequency of the pump field.

We should note that the results presented here are strictly specific to the degenerate manifolds of the $F_g= 1 \rightarrow F_e=2$ hyperfine transition. 
For the system made of $F_g=2$ and $F_e=3$ states, the contribution from the population inverted states into the amplification of spontaneous emission may be more significant.

\subsection{Propagation Results}
\begin{figure}[h!]
    \centering
    \includegraphics[scale=0.5]{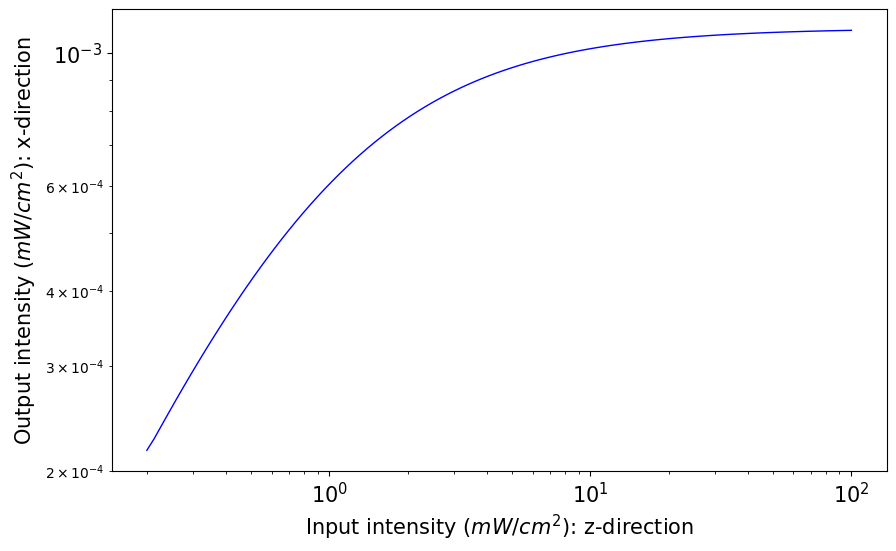}
    \caption{The output intensity of the field linearly polarized in the $x$ direction as a function of the pump field linearly polarized in the z-direction. The cell length is $0.1\,$m and the atomic density is $n=1.16\times10^{16}\,$m$^{-3}$ \cite{Papoyan2019}.} \label{ASE_Input_vs_Output}
\end{figure}

In this section, we solve Eq.\,\eqref{propagation_equations} to find the output  intensity of the field having linear polarization orthogonal to polarization of the incident pump field. Once the system is pumped to a steady state, the spontaneous photons generate light, which further gets amplified or absorbed depending on the contribution from the first term of Eq.\,\eqref{propagation_equations}. 
The output  field intensity for varying values of the pump field intensity is given in Fig.\,\ref{ASE_Input_vs_Output}. The output intensity  initially varies linearly and is followed by saturation. The generated intensity is built from  two contributions: one coming from the directional, stimulated process due to induced polarization [first term in Eq.\,\eqref{propagation_equations}] and the second coming from the amplification of isotropic spontaneous emission [second term in Eq.\,\eqref{propagation_equations}]. Population inversion present in the steady atomic state contributes to the stimulated emission/absorption mechanism 
yet is not enough to overcome overall losses (absorption from non-inverted states) and hence does not lead to gain but only reduced absorption. 
However, one should expect the output measurement of the field  having polarization perpendicular to the pump field to be nonzero owing to the isotropic emission of spontaneously emitted photons.
The intensity of this field depends on the Rabi frequency and the pump field detuning.  
As an example, assuming a decay rate $\Gamma= 2\pi \times 5.6$\,MHz ($^{87}$Rb $F=2\rightarrow F=3$),  we propose using a Rabi frequency  $\Omega_p=0.4\Gamma$, and a one-photon detuning $\Delta_p=0.75\Gamma$ for experimental realization.

\section{Experimental situation}\label{sec:experiments}

The idea of degenerate mirrorless lasing was born at the Institute for Physical Research, in Ashtarak, Armenia \cite{movsisyan,gazazyan} and evolved into a fruitful collaboration with the University of Mainz a few years later. 



The principle idea, as discussed above, is that optical pumping redistributes the population among the ground-state Zeeman sublevels. At sufficiently high light powers of a linearly-polarized radiation, the population is partially transferred to the excited state, which, in the case of an $F\rightarrow F+1$ transition may result in population inversions between certain sublevels of the ground and excited states. Due to the pencil-shaped geometry of the light-atom interaction volume, this inversion may, under right circumstances, result in directional emission from the sample.
In the case of degenerate mirrorless lasing, amplified spontaneous emission is produced along or opposite to the pump-beam direction, with polarization orthogonal to that of the pump. 



Forward degenerate mirrorless lasing was detected both in Ashtarak and Mainz, showing a typical threshold dependence of the outgoing beam on the incoming beam intensity. The transitions for which we expected and observed the lasing are $F_{e} > F_{g}$ cycling transitions in the $D_2$ line of rubidium, specifically, $^{85}$Rb $F_{g}=3 \rightarrow F_e=4$ and $^{87}$Rb $F_g=2 \rightarrow F_e=3$. The intensity of the emission polarized orthogonally with respect to the incident light was studied as a function of the pump intensity and magnetic field applied along the direction of the pump-light polarization. Up to 1\,\% conversion efficiency was observed at zero-magnetic field, which approached to zero with the applied magnetic field. The width of this feature was about 15 -- 100 mG (FWHM) depending on the pump intensity. 

These findings \cite{Papoyan2019} are a strong indication of amplified emission in the forward direction, also known as \textit{forward mirrorless lasing}. In parallel, we detected degenerate four-wave mixing (FWM), which makes the signal more complicated.

While forward mirrorless lasing is an interesting phenomenon, backward mirrorless lasing is of greater interest, since it can be used for remote-detection magnetometry, for example with laser guide stars \cite{bustos}. Indeed, high directivity of the backward beam will significantly facilitate recording of the return flux, which will make the distant remote sensing much easier. 

From the theoretical analysis presented in previous sections one can expect that the mirrorless lasing should also take place along the pump beam in the backward direction. 

It would seem that mirrorless lasing in the backward direction should be detected even more easily than in the forward direction, since there is no strong background of pump radiation, and also no contribution from the accompanying FWM. However, preliminary experiments have revealed complexities that require further reflection and clarification.

It was possible to record a non-diverging backward emission in Ashtarak with much lower conversion efficiency (almost 3 orders of magnitude less than for the forward one) at specific experimental conditions. The recorded radiation had an intensity threshold and was sensitive to the applied magnetic field, similar to the forward case, but exhibited sharp sub-Doppler features, which were not observed in the forward lasing. But this result could not be reproduced in Mainz, even with the same vapor cell and the same experimentalists.

However, these preliminary studies were useful in identifying the factors important for realization of the backward mirrorless lasing. Among these factors are: (i) careful avoiding of residual birefringence of the vapor cell windows; (ii) proper cancellation and control of the magnetic field; (iii) presence (or absence) of a weak seeding radiation; (iv) broad spectral linewidth of the pump laser. 

The much lower conversion efficiency, as well as the narrow spectral features of the backward radiation, should be associated with the competition between the emission and absorption in opposite directions by atoms with different velocities for different pump detunings from Doppler-overlapped atomic transitions. 

Further theoretical analysis and experimental studies are needed for complete understanding, as well as unambiguous and reproducible demonstration of backward degenerate mirrorless lasing.



\section{Conclusions and outlook}

Degenerate mirrorless lasing is a beautiful phenomenon that has been realized experimentally in the forward direction. However, in the case of backward directed emission, more experimental work needs to be done. At the same time, theoretical guidance is needed to guide and interpret the experimental work.

In this manuscript, we have reviewed the current status of the mirrorless lasing and presented the work-in-progress developments in the general theoretical analysis, hopefully, on the way to a comprehensive solution of the problem.

\clearpage

\section*{Conflicts of Interest}
The authors declare no conflicts of interest.

\begin{acknowledgments}

We recognize the contribution of the tragically deceased Hrayr Azizbekyan.
S.M., J.Ch., and A.R. acknowledge support from the Office of Naval Research under awards N00014-20-1-2086 and N00014-22-1-2374. S.M. acknowledges the Helmholtz Institute Mainz Visitor Program and the Alexander von Humboldt Foundation. J.Ch. was also partially supported by DEVCOM Army Research
Laboratory under Cooperative Agreement Number
W911NF2320076. D.B. was partially supported by DEVCOM Army Research Laboratory under Cooperative Agreement Number W911NF2120180.
\end{acknowledgments}

\clearpage
\appendix
\section{Spontaneous emission modes}
We use the formalism from \cite{Agarwal1974}:

We define the field operator:

\begin{equation}
    E^{(\mu)+}_H(\vec{r},t)=\dfrac{i}{(2\pi)^3}\int d^3k\text{ }{a}_{\vec{k},H}^{(\mu)}(t)\sqrt{\dfrac{h\omega_k}{2\epsilon_0}}e^{i\vec{k}\cdot\vec{r}}\varepsilon^{(\mu)}_{\vec{k}}\,,
\end{equation}

and use the Heisenberg equations of $a_{\vec{k}}^{(\mu)}(t)$:

\begin{equation}
    \dot{a}_{\vec{k},H}^{(\mu)}(t)=-i\omega_k{a}_{\vec{k},H}^{(\mu)}(t)-i\sum_{\alpha,j}g_{\vec{k},j}^{(\mu)*}(\vec{R}_{\alpha})\sigma^{+}_{j,H}(\vec{R}_{\alpha},t)\,,
\end{equation}

to get the following equation for the Heisenberg operator:
\begin{equation}
\begin{split}
    &\Box^2E^{(\mu)+}_H(\vec{r},t)=\dfrac{1}{(2\pi)^{3/2}c}\sum_{\alpha,j}\int d^3k\text{ }\sqrt{\dfrac{h\omega_k}{2\epsilon_0}}g_{\vec{k},j}^{(\mu)*}(\vec{R}_{\alpha})\varepsilon^{(\mu)}_{\vec{k}}\\
    &\cross(\omega_k-\omega_j)\sigma^{+}_{j,H}\left(\vec{R}_{\alpha},t-\dfrac{\abs{\vec{r}-\vec{r}^{\prime}}}{c}\right)e^{i\vec{k}\cdot\vec{r}}.
\end{split}
\end{equation}

$\Box^2$ is the D'Alembertian. The right hand side is the total current density, summing over all atomic contributions.
We use the Green's function for the D'Alembertian to get the integral:

\begin{equation}
\begin{split}
    &E^{(\mu)+}_{j,H}(\vec{r},t)=\dfrac{1}{c}\int d^3r^{\prime}\text{ }\dfrac{J_j^+\left(\vec{r}^{\prime},t-\dfrac{\abs{\vec{r}-\vec{r}^{\prime}}}{c}\right)}{\abs{\vec{r}-\vec{r}^{\prime}}}\,,
\end{split}
\end{equation}

and obtain the general solution:

\begin{equation}
\begin{split}
    &E^{(\mu)+}_{j,H}(\vec{r},t)=\sqrt{\dfrac{h}{2c^2\epsilon_0(2\pi)^3}}\sum_{\alpha,j}\int d^3r^{\prime}\int dk d\Omega_k\\
    &\cross g_{\vec{k},j}^{(\mu)*}\left(\vec{R}_{\alpha}\right)k^3(\omega_k-\omega_j)\dfrac{\hat{k}\cross(\hat{k}\cross\hat{d}_j)}{\abs{\vec{r}-\vec{r}^{\prime}}}\\
    &\cross \sigma^{+}_{j,H}\left(\vec{R}_{\alpha},t-\dfrac{\abs{\vec{r}-\vec{r}^{\prime}}}{c}\right)e^{i\vec{k}\cdot\vec{r}}.
\end{split}
\end{equation}

In the far-field limit, the solution simplifies to:

\begin{equation}
\begin{split}
    &E^{(\mu)+}_{j,H}(\vec{r},t)\propto \sum_{\alpha,j}\dfrac{\omega_j^2}{c^2}\dfrac{(\hat{r}_{\alpha}\cdot\hat{d}_j)\hat{r}_{\alpha}-\hat{d}_j}{r_{\alpha}}\\
    &\cross\sigma^{+}_{j,H}\left(\vec{R}_{\alpha},t-\dfrac{r}{c}\right)e^{-i\omega_j(\hat{r}\cdot\hat{r}_j)}.
\end{split}
\end{equation}

where $\vec{r}_{\alpha}=\vec{r}-\vec{R}_{\alpha}$.

In the far field, each atom acts as a dipole antenna. In the case where we have many atoms and they're all stationary, a very complicated spatially constant interference pattern emerges. However if the atom has a statistical velocity distribution and the number density is high enough for collisions to be significant, the atomic medium behaves as a choatic light source.

\section{Paraxial approximation}
The volume-averaged dipole operator $S_{j;I^{\prime}}^{+}(\vec{R}_{i},t)=n(\vec{R}_{i})\int_{V_r(\vec{R}_i)}dV^{'} \sigma_{j;I^{\prime}}^{+}(\vec{R}^{'},t)$ is introduced. The volume $V_r(\vec{R}_i)$, center at $\vec{R}_i)$ with radius $r$  is chosen as our sampling region where the atomic density matrix and the radiation fields vary trivially in $V_r(\vec{R}_i)$. We use it to simplify the below quantity:

\begin{equation}
\begin{split}
    &\int_{V_r(\vec{R}_i)}dV^{'}\mathcal{C}^{(\mu^{\prime})}_{j,l}(\vec{r},\vec{R}^{'},t,t^{\prime})e^{i(\vec{q}-\vec{k}^{\prime})\cdot\vec{R}}\\
    &=\mathcal{C}^{(\mu^{\prime})}_{j,l}(\vec{r},\vec{R}_i,t,t^{\prime})\int_{V_r(\vec{R}_i)}dV^{'}e^{i(\vec{q}-\vec{k}^{\prime})\cdot\vec{R}}.
\end{split}
\end{equation}

In the limit of high atomic number density, the condition $\vec{q}=\vec{k}^{\prime}$ is enforced as other contributions average to 0 as only contributions where the wavevectors are approximately the same survive phase averaging over the volume. With the paraxial approximation, the dot product $\varepsilon^{(\mu)}_{\vec{k}}\cdot\varepsilon^{(\mu^{\prime})}_{\vec{k^{\prime}}}\approx\delta_{(\mu\mu^{\prime})}$. We assume the averaged dipole moment vector $\vec{D}_l(\vec{R}_i)=\int_{V_r(\vec{R}_i)}dV^{'}\vec{d}_l(\vec{R}^{\prime})$ averages out all fluctuations such that the z-dipole moment, $\vec{D}_z(\vec{R}_i)$, is oriented along the pump polarization axis. Ignoring the Langevin terms, Eq. \eqref{SnExp} simplifies to:

\begin{align}\label{SnExpR1}
    \begin{split}
        &\mathcal{R}_{\text{s}}(t)\\
        &=-\delta_{(\mu\mu^{\prime})}\sum_{jl,\alpha;\nu=\pm}\int_0^t dt^{\prime}\\
        &\left(\expval{(E_{I^{\prime}}^{(\mu)-}(\vec{r},t) \vec{E}_{I^{\prime}}^{(\mu)+}(\vec{r},t^{\prime})(\vec{d}_l\cdot\varepsilon^{(\nu)}_{\vec{x}})\mathcal{D}_{j,l}(\vec{R}_{i},t,t^{\prime})}\right.\\        &\left.+\expval{\vec{E}_{I^{\prime}}^{(\mu)-}(\vec{r},t^{\prime})E_{I^{\prime}}^{(\mu)+}(\vec{r},t)(\vec{d}^{*}_l\cdot\varepsilon^{(\nu)*}_{\vec{x}})\mathcal{D}^{\dag}_{j,l}(\vec{R}_{i},t,t^{\prime})}\right)/,,
    \end{split}
\end{align}

Where $\mathcal{D}_{j,l}(\vec{R}_{i},t,t^{\prime})$ is the dipole correlation function Eq. \eqref{Dipcor}.

The expectation of single atom dipole products $\expval{{\sigma}^{+}_{l,I^{\prime}}(\vec{R_i},t){\sigma}^{-}_{m,I^{\prime}}(\vec{R_i},t^{\prime})}$ is expanded in the dressed state basis in terms of operators, defined in the picture $I^{\prime}$, ${\sigma}_{\Lambda_{r}\Lambda_{r^{\prime}}}(\vec{R_i},t,t^{\prime})=\ket{\Lambda_{r}(\vec{R_i},t)}\bra{\Lambda_{r^{\prime}}(\vec{R_i},t^{\prime})}$. :

\begin{equation}
\begin{split}
    &\sigma^{+}_{i,I^{\prime}}(\vec{R_{\alpha}},t){\sigma}^{-}_{j,I^{\prime}}(\vec{R_{\alpha}},t^{\prime})\\
    &=-{i}{\hbar}\sum_{i,l,l^{\prime}}\sum_{j,m}C^{i}_{ll^{\prime}}(\vec{R}_{\alpha},t)C^{j*}_{ml^{\prime}}(\vec{R}_{\alpha},t^{\prime})e^{it(\omega_i-\omega_j+\lambda_{l}-\lambda_m)}\\
    &\cross\sigma_{\Lambda_l\Lambda_m}(\vec{R}_{\alpha},t,t^{\prime})e^{i(\omega_j+\lambda_m-\lambda_{l^{\prime}})(t-t^{\prime})}/,,
\end{split}
\end{equation}

Assuming the weak coupling approximation, the Markov approximation along with the slow change of $\sigma_{\Lambda_l\Lambda_m}(\vec{R}_{\alpha},t,t^{\prime})$, the coefficients $C_{mm^{\prime}}(t^{\prime})$ and the operators are dependent only on time $t$. The time evolution of the expectation of these operators is given below (The contribution of the stimulated emission and pair-atom terms are ignored):

\begin{equation}
\begin{split}
    &\dfrac{d}{dt}\expval{\sigma_{\Lambda_a\Lambda_b}(\vec{R}_{\alpha},t,t)}\\
    &=-\dfrac{i}{\hbar}\int\int d^3k\text{ }d^3k^{\prime}\text{ }g_{\vec{k},i}^{(\mu)}(\vec{R}_{\alpha})g_{\vec{k}^{\prime},j}^{(\mu^{\prime})*}(\vec{R}_{\alpha})\\
    &\cross \sum_{lm}C^{i}_{la}(\vec{R}_{\alpha},t)C^{j*}_{am}(\vec{R}_{\alpha},t)e^{it(\omega_i+\lambda_{l}-\omega_k)}\\
    &\cross\int_{t_0}^{t}dt^{\prime}\expval{\sigma_{\Lambda_a\Lambda_m}(\vec{R}_{\alpha},t,t)}e^{-it^{\prime}(\omega_j+\lambda_{m}-\omega_{k^{\prime}})}/,,
\end{split}
\end{equation}

We now use the Markov approximation. 
From the standard methods of solving for the spectral line shapes by solving the above equations for all dipole operators in Laplace space, we can determine the power broadened spectral line shapes, $f_{\Lambda_a\Lambda_m}(\omega^{\prime},t)$, corresponding to $\expval{\sigma_{\Lambda_a\Lambda_m}(\vec{R}_{\alpha},t,t)}$ and the decay rate $\Gamma_{\Lambda_a\Lambda_m}(t)$. The spontaneous decay rates for the atomic density matrices in the dressed state basis are given below:

\begin{equation}
\begin{split}
    &\Gamma_{1}^{ab;ml}(t)=\int_{0}^{\infty}d\tau\int d^3k\sum_{ij}\left(C_{bl}^{i*}(t)C_{ma}^{j}(t)g_{\vec{k},i}^{(\mu)*}g_{\vec{k},j}^{(\mu)}\right.\\
    &\left.e^{-i(\omega_{ij}+\lambda_{bl}+\lambda_{am})t}e^{i(\omega_k-\omega_j-\lambda_{am})\tau}\right.\\
    &+\left.C_{ma}^{i}(t)C_{bl}^{j*}(t)g_{\vec{k},i}^{(\mu)}g_{\vec{k},j}^{(\mu)*}e^{i(\omega_{ij}+\lambda_{bl}+\lambda_{am})t}e^{-i(\omega_k-\omega_j-\lambda_{bl})\tau}\right)/,,
\end{split}
\end{equation}

\begin{equation}
\begin{split}
    &\Gamma_{2}^{ab;l}(t)=\int_{0}^{\infty}d\tau\int d^3k\sum_{ijl^{\prime}}\left(C_{l^{\prime}l}^{i*}(t)C_{l^{\prime}b}^{j}(t)g_{\vec{k},i}^{(\mu)*}g_{\vec{k},j}^{(\mu)}\right.\\
    &\left.e^{-i(\omega_{ij}+\lambda_{l^{\prime}l}+\lambda_{bl^{\prime}})t}e^{i(\omega_k-\omega_j-\lambda_{bl^{\prime}})\tau}\right)/,,
\end{split}
\end{equation}

\begin{equation}
\begin{split}
    &\Gamma_{3}^{ab;l}(t)=\int_{0}^{\infty}d\tau\int d^3kd^3k^{\prime}\sum_{ijl^{\prime}}\left(C_{ll^{\prime}}^{i}(t)C_{l^{\prime}a}^{j*}(t)g_{\vec{k},i}^{(\mu)}g_{\vec{k},j}^{(\mu)*}\right.\\
    &\left. e^{i(\omega_{ij}+\lambda_{l^{\prime}a}+\lambda_{l^{\prime}l})t}e^{-i(\omega_k-\omega_j-\lambda_{l^{\prime}a})\tau}\right)/,,
\end{split}
\end{equation}

\bibliographystyle{unsrt}
\bibliography{mirrorlessbib}

\end{document}